%% file: spin_paper.tex
\documentclass[twocolumn,aps,prd,floatfix,preprintnumbers,a4paper,nofootinbib,
superscriptaddress,10pt]{revtex4-1}

\usepackage{epsfig,graphics,graphicx}
\usepackage{amsmath,amssymb,mathtools,amsfonts}
\usepackage{bm,soul}
\usepackage[usenames,dvipsnames]{color}
\usepackage{amssymb,url,psfrag,times}
\usepackage[varg]{txfonts}
\usepackage[colorlinks, pdfborder={0 0 0}]{hyperref}
\usepackage{lineno,verbatim}
\usepackage[utf8]{inputenc}
\usepackage{ulem}
\usepackage{xcolor}
\usepackage[nolist,nohyperlinks]{acronym}

\usepackage[english]{babel}
\usepackage{blindtext}

\definecolor{LinkColor}{HTML}{60a567}
\definecolor{CiteColor}{HTML}{60a567}
\definecolor{UrlColor}{HTML}{60a567}
\hypersetup{linkcolor=LinkColor}
\hypersetup{citecolor=CiteColor}
\hypersetup{urlcolor=UrlColor}
\input{misc-definitions}

\def\rd{ringdown}

\def\hf{\tilde{h}}
\def\h22{\hf_{22}}

\def\A22{A_{22}}
\def\vphi22{\varphi_{22}}

\def\f22{f_{22}}

\begin{document}

\newcommand{\Cardiff}{School of Physics and Astronomy, Cardiff}
\newcommand{\Sapienza}{Dipartimento di Fisica, Universit\`{a} di Roma ``Sapienza'', Piazzale A. Moro 5, I-00185, Roma, Italy}
\newcommand{\INFN}{INFN Sezione di Roma, Piazzale A. Moro 5, I-00185, Roma, Italy}


\title{Investigating the effect of in-plane spin directions for Precessing BBH systems. }

\author{Chinmay Kalaghatgi}
\affiliation{\Cardiff}

\author{Mark Hannam}
\affiliation{\Cardiff}
\affiliation{\Sapienza}
\affiliation{\INFN}


\begin{abstract}

In gravitational-wave observations of binary black holes (BBHs), theoretical waveform models are used to infer the black-hole properties. 
There are several sources of potential systematic errors in these measurements, including due to physical approximations in the models. 
One standard approximation is to neglect a small asymmetry between the $+m$ and $-m$ spherical-harmonic modes; this is the effect
that leads to emission of linear momentum perpendicular to the orbital plane, and can result in large recoils of the final black hole. 
The asymmetry is determined by both the magnitude \emph{and} direction of the spin components that lie in the orbital plane. We investigate the 
validity of this approximation by comparing numerical relativity (NR) simulations of single-spin NR systems with varying in-plane spin 
directions and magnitudes (including several ``superkick'' configurations). We find that the mode asymmetry will impact measurements
at signal-to-noise ratios (SNRs) between 15 and 80, which is well within current observations. In particular, mode asymmetries are
likely to impact measurements at comparable SNRs to those at which we might hope to make the first unambiguous measurements of 
orbital precession. We therefore expect that models will need to include mode-asymmetry effects to make unbiassed 
precession measurements.

\end{abstract}

\date{\today}

\maketitle


\begin{acronym}
\acrodef{GW}[GW]{gravitational-wave}
\end{acronym}

\section{Introduction}
Since the advent of the Advanced LIGO~\cite{TheLIGOScientific:2014jea} and Virgo~\cite{TheVirgo:2014hva} gravitational-wave 
detectors in 2015, up to October 2019 there have been a total of 50 gravitational wave signals detected which include multiple binary-black hole mergers, binary neutron star mergers and possible neutron-star-black-hole merger~\cite{LIGOScientific:2018mvr, Abbott:2020niy, Nitz_2020, Zackay:2019btq, Venumadhav:2019lyq, Abbott:2020khf, LIGOScientific:2020stg,Abbott:2020uma, Abbott:2020tfl}. 
The BBH observations have begun to reveal the astrophysical rate of black-hole mergers, and the astrophysical distribution of 
black-hole masses and spins~\cite{LIGOScientific:2018jsj, Abbott:2020mjq, Abbott:2020gyp}. To measure the binary's properties the detector data are 
compared against a set of theoretical model waveforms. 
The accuracy of the measured parameters depends not only on the details of the source, the signal-to-noise ratio (SNR) of the signal, 
and parameter degeneracies, but also on the accuracy of the waveform models. Two families of waveform models, 
\texttt{IMRPhenom}~\cite{Khan:2015jqa,London:2017bcn,Hannam:2013oca,Khan:2018fmp,Pratten:2020fqn,Pratten:2020ceb} and 
\texttt{SEOBNR}~\cite{Bohe:2016gbl,Cotesta:2018fcv,Pan:2013rra,Babak:2016tgq,Taracchini:2013rva}, were used to calculate the 
reported parameters during the first two observation runs. Both rely on several physical approximations, as discussed in, for 
example, Ref.~\cite{Ramos-Buades:2020noq}. In this paper we test the 
validity and impact of a subset of those approximations. 

Binary-black-hole inspiral is the result of orbital energy and angular momentum loss through gravitational radiation. 
If the radiation from a binary is decomposed into spin-weighted spherical harmonics, $^{-2}Y_{\ell m}(\theta,\phi)$, 
the signal is dominated by the ``quadrupole'' contribution in the $\ell=2$ harmonics. Gravitational waves 
also carry linear momentum, and for nonspinning or aligned-spin binaries (where the black-hole spins $\mathbf{S}_i$ are 
parallel to the orbital angular momentum, $\mathbf{L}$, so that $\mathbf{L} \times \mathbf{S}_i = 0$), the resultant recoil of 
the center-of-mass within the orbital plane is manifest in the signal through interplay between different multipoles; see, for 
example, Ref.~\cite{Herrmann:2007ac}.
Current aligned-spin binary waveform models that include higher multipoles capture all of these physical effects, with varying 
degrees of accuracy~\cite{London:2017bcn, Garcia-Quiros:2020qpx, Cotesta:2018fcv}. In generic binaries, where the spins are mis-aligned with the orbital angular momentum, the orbital plane and spins precess
during the inspiral. Generic binaries also radiate linear momentum perpendicular to the orbital plane. This effect, which shows up in the 
GW signal through an asymmetry between the $+m$ and $-m$ multipoles, is not present in current precessing \texttt{SEOBNR}~\cite{Ossokine:2020kjp} and \texttt{IMRPhenom}~\cite{Khan:2019kot} models. Although these models include the spin directions while computing the precession dynamics used to generate the precessing waveform, the effect of varying spin directions on the full waveforms remains un-modelled. Our goal is to make a first estimate of the effect of these omissions on GW source parameter measurements. 

We begin by describing in more detail the phenomenology of BBH systems, and the construction of generic-binary waveform models. 

A BBH system undergoing non-eccentric inspiral can be characterised by eight parameters, the individual masses $(m_{i})$, and 
the components of the two spin vectors $(\mathbf{S}_{i})$, specified at some fiducial point during the inspiral, for example a chosen
orbital frequency. The GW signal is also parameterised by the binary's sky-position $(\alpha, \delta)$, inclination $(\iota)$, 
 coalescence phase ($\phi_c$), distance $(d_{L})$, polarisation $(\psi)$ and time of arrival $(t_c)$ at the detector. 
As noted above, the complex GW strain can be decomposed into spin-weighed spherical harmonics as, 
\begin{equation}~\label{eq:strain_decomposed}
h(t, \theta, \phi) = h_{+}(t) - i h_{\times}(t) = \sum_{\ell,m} h_{\ell m}(t) \, ^{-2}Y_{\ell m}(\theta, \phi),
\end{equation}
where $(\theta, \phi)$ give the position of the observer on a sphere centred on the centre-of-mass of the binary. 

Based on the black-hole (BH) spin configurations, coalescing BBH systems with spins can be considered to be either:

	\begin{itemize}
	
	\item \textbf{Aligned-Spin:} The BH spins are parallel or anti-parallel to $\mathbf{L}$, so $\mathbf{L} \times \mathbf{S}_i = 0$, where 
$i=1,2$ for each BH. From the symmetries of the system, the BHs orbit in a fixed plane, i.e., the direction of the orbital angular 
momentum $\mathbf{\hat{L}}$ remains fixed. In the frame where $\mathbf{\hat{L}} \parallel \hat{z}$, symmetry also implies 
that \begin{equation}
h_{\ell,m} = (-1)^{\ell}h_{\ell, -m}^{*}, \label{eqn:symmetry}
\end{equation}
and that any linear momentum emission is perpendicular to $\mathbf{L}$; although the 
orientation of the orbital plane remains fixed, the center-of-mass can recoil within this plane.  

	\item \textbf{Precessing:} One or both BHs have non-zero spin components perpendicular to $\mathbf{\hat{L}}$. 
We denote the parallel components by $\mathbf{S}_i^{\parallel}$ and the perpendicular components by $\mathbf{S}_i^{\perp}$. 
The presence of $\mathbf{S}_{i}^{\perp}$ causes the 
orbital plane to precess over the course of the coalescence. This leads to modulations of the 
amplitude and phase of the waveform. Emission of linear momentum is now also possible perpendicular to the orbital plane, which 
breaks the symmetry of Eq.~(\ref{eqn:symmetry}) between the $\pm m$ multipoles. 

	\end{itemize}

As was shown in previous studies~\cite{Schmidt:2012rh,Boyle:2011gg,OShaughnessy:2011pmr}, a precessing waveform can be 
decomposed into the waveform as observed in a co-precessing frame, $h^{\rm{CP}}$, and a time- or frequency-dependent rotation
that describes the precessional dynamics. The rotation can be expressed in terms of three Euler angles, $(\alpha, \beta, \gamma)$, 
and the $\ell=2$ modes of the precessing-binary waveform $h^{\rm P}$ constructed as \begin{equation}~\label{eq:wigner_rotation}
h^{\rm P}_{2m} = e^{im\alpha} \sum_{m'} e^{- i m' \epsilon} d^{2}_{m m'}(\beta) \, h_{2 m'}^{\rm{CP}},
\end{equation}
where $d^2_{m m'}$ denote the $\ell = 2$ Wigner-d matrices. 
In the current precessing models (\texttt{IMRPhenom} and \texttt{SEOBNR}), the co-precessing-frame waveform is based on an 
underlying non-precessing-binary model (with some modifications), and this procedure preserves its orbital-plane symmetry, 
Eq.~(\ref{eqn:symmetry}). These models therefore do not include the $\pm m$ mode asymmetry of full precessing-binary waveforms. 

The magnitude and direction of the out-of-plane angular momentum loss $\dot{p}_{\parallel}$ 
(and therefore the level of mode asymmetry) is related to the angles between the in-plane spins $\mathbf{S}_i^{\perp}$ and the 
separation vector between the two black holes $\mathbf{\hat{n}}$, as most easily seen in the PN treatment in Sec.~III.E of Ref.~\cite{PhysRevD.52.821}. 
During one orbit the spin directions change little, so $\dot{p}_{\parallel}$ oscillates approximately on the orbital timescale. In the
``twisted-up'' models described above, this effect is not present, and an overall rotation of the spin(s) in the orbital plane introduces 
only an offset in the precession angle $\alpha$, which is degenerate with the azimuthal angle, $\phi$, since it enters the spin-weighted
spherical harmonics as $e^{i m \phi}$. The model waveforms are therefore degenerate with respect to a constant rotation of the 
in-plane spins, while true waveforms include an additional effect that varies sinusoidally with respect to this spin rotation.

Out of plane recoil in the context of mode asymmetries has been discussed in NR simulations in~\cite{Brugmann:2007zj}, and further illustration of the effect in GW signals is shown in~\cite{Ramos-Buades:2020noq}. Earlier studies on in-plane effects on waveforms and/or mode-asymmetries for precessing systems include~\cite{OShaughnessy:2012iol, Pekowsky:2013ska, Boyle:2014ioa}.

In this study, we investigate the effects of varying the in-plane spin direction for single-spin precessing NR waveforms for a 
given combination of  mass-ratio and spin. 
We also consider the special case of the ``super-kick'' configuration~\cite{Campanelli:2007cga,Gonzalez:2007hi,Brugmann:2007zj}: these
are equal-spin configurations where the spins lie entirely in the orbital plane, and $\mathbf{S}_1 = - \mathbf{S}_2$. Due to the 
symmetry of this configuration, the orbital plane does not precess, but \emph{does} bob up and down due to linear momentum
loss, making this an especially clean system for the study of mode asymmetry. 
 We choose these configurations to estimate the importance of mode-asymmetric content on parameter measurements. 
Using the waveform with in-plane spin initially aligned to the position vector as a \emph{proxy template}, we compute matches 
(see Sec.~\ref{sec:match_computation}) against systems 
with different in-plane spin directions. Using a relationship between the match value and SNR at which two signals are distinguishable, 
we provide an estimate of the SNR at which mode asymmetries will impact parameter measurements. We also use a selection of 
waveforms with the same spin direction as the proxy template but with differing in-plane spin magnitude to estimate the relative strength of the 
effect of varying spin direction versus varying spin magnitude. 

For all results, we use only the $(\ell = 2, m=\pm2)$ modes of the waveforms in the co-precessing frame. Higher modes are much weaker
than the $(\ell=2, |m|=2)$ multipoles, but far stronger than the asymmetry contribution to the dominant modes, and we choose to 
consider only the dominant modes in order to more easily isolate effects due to the mode asymmetry.



The paper is organised as follows. Sec.~\ref{sec:simulation_details} provides details of the simulations generated for this study,
Sec.~\ref{sec:match_computation} and Sec.~\ref{sec:match_connection} discuss the computation of precessing matches and 
the connection between the match and detectable SNR respectively. The specific results presented are motivated in Sec.~\ref{sec:setup} 
with the actual results in Sec.~\ref{sec:results}. The conclusions we draw from this work, and some of its limitations and potential future
extensions, are discussed in Sec.~\ref{sec:conclusions}.

\section{NR Waveforms}~\label{sec:simulation_details}

	\begin{table*}[t]
	\begin{tabular}{|c|c|c|c|c|c|c|c|c|}
	\hline 
	Config & q & $\vec{S}_2$ & $\vec{r} = D/M$ & $\vec{p} = \vec{p_1} - \vec{p_2}$ & $\omega_{start}(fM)$ & $\phi_{\rm{Sn}}$ & $\theta_{SL}$ \\
	\hline
	q1a08p0$_{\rm{sk}}$ & 1 & (0, -0.799, -0.001) & ( 0, 11.623, 0)&(-0.174, -0.001, 0)& 0.0225 & 0  & $\pi/2$  \\
	q1a08p90$_{\rm{sk}}$ & 1 & (0.7999, 0, -0.0012)  & ( 0, 11.623, 0)&(-0.174, -0.001, 0)  & 0.0225 & $\pi/2$ & $\pi/2$  \\
	q1a08p180$_{\rm{sk}}$ & 1 & (0, 0.7999, -0.0012)  & ( 0, 11.623, 0)&(-0.174, -0.001, 0)& 0.0225  & $\pi$ & $\pi/2$ \\
	q1a08p270$_{\rm{sk}}$ & 1 & (-0.7999, 0, -0.0012)  & ( 0, 11.623, 0) &(-0.174, -0.001, 0)& 0.0225 & $3 \pi/2$ & $\pi/2$  \\
	\hline
	\hline
	q2a07p0 & 2 & (-0.001, 0.699, 0.006)  & ( 0., 10.810,   0. ) & (-0.105, -0.001, 0.123) & 0.025 & 0  & $\pi/2$  \\
	q2a07p90 & 2 & (-0.451, -0.005,  0.535) & ( 0., 10.810,   0. ) & (-0.105, -0.001, 0.123)& 0.025& $\pi/2$ & $\pi/2$  \\
	q2a07p180 & 2 & (0.006, -0.699, -0.002)  & ( 0., 10.810,   0. )& (-0.105, -0.001, 0.123)& 0.025 & $\pi$ & $\pi/2$ \\
	q2a07p270 & 2 & (0.448, -0.005, -0.537)  & ( 0., 10.810,   0.  )& (-0.105, -0.001, 0.123)& 0.025 & $3 \pi/2$ & $\pi/2$  \\
	\hline
	\hline
	q4a08p0 & 4 & (0.0007, 0.799, -0.005) & ( 0.     , 11.486,   0.  )&(-0.111, -0.0004, 0.014)& 0.0225 & 0  & $\pi/2$  \\
	q4a08p90 & 4 & (-0.793, 0, 0.099) & ( 0.     , 11.486,   0.  )&(-0.111, -0.0005, 0.014)  & 0.0225 & $\pi/2$ & $\pi/2$  \\
	q4a08p180 & 4 & (-0.0007, -0.799, -0.005)  & ( 0.     , 11.486,   0.  )&(-0.111, -0.0004, 0.014)& 0.0225  & $\pi$ & $\pi/2$ \\
	q4a08p270 & 4 & (0.792, 0, -0.110)  & ( 0.     , 11.486,   0.  ) &(-0.111, -0.0005, 0.0147)& 0.0225 & $3 \pi/2$ & $\pi/2$  \\
	\hline
	\hline
	q4a04p0 & 4 & (-0.001, 0.399, -0.00007)  & ( 0.     , 11.486,   0.  )&(-0.111, -0.0004, 0.014)& 0.0299 & 0  & $\pi/2$  \\
	q2a04p0 & 2 & (-0.00008, 0.3999, -0.0008) & ( 0.     , 11.6299,   0.  )&(-0.153, -0.0009, 0.015)& 0.0224 & 0  & $\pi/2$  \\
	q2a08p0 & 2 & (0.0005, 0.799, -0.003)  & ( 0.     , 11.5709,   0.  )&(-0.153, -0.0009, -0.0243)& 0.023 & 0  & $\pi/2$  \\
	\hline
	\end{tabular}
	\caption{ Table of NR simulations used for this study. From left to right, the columns show the name of the simulation, 
	the mass-ratio of the system, value of the spin on the larger black hole at the reference frequency, the separation between 
	the black holes at the reference frequency, the total momenta of the system at the reference frequency, the reference frequency 
	at which the simulation starts, and the values of the $\phi_{\rm{Sn}}$ and $\theta_{SL}$ angles respectively. For the $q= 1$ series, 
	note that $\mathbf{S}_{2} = - \mathbf{S}_{1}$. }\label{Tab:nr_waveforms_list_ch4}
\end{table*}

For this study, a set of 12 new NR simulations were performed with the BAM code~\cite{Husa_2008,PhysRevD.77.024027}.
Configurations are defined by the mass ratio, $q = m_2/m_1$, where we choose the convention $m_2 > m_1$, and the spin
vectors specified at the start of the simulation, $\mathbf{S}_i$. In the unequal-mass simulations, only the larger black hole is spinning, 
so that $\mathbf{S}_1 = 0$. (We could also assign spin to the secondary black hole, but placing spin on one black hole is 
sufficient to produce the asymmetry effects that we wish to study.) For these simulations we can completely specify the spin 
direction at the beginning of the simulation by two angles, i) the angle between the spin vector and angular momentum vector,
which we call $\theta_{SL}$, and ii) the angle between the separation vector ($\vec{n}$) and the projection of spin onto the orbital plane ($\mathbf{S}_i^{\perp}$), 
which we call $\phi_{\rm{Sn}}$. The codes available at the beginning of this study for initial data generation did not allow for user 
specified $(\theta_{SL}, \phi_{\rm{Sn}})$ values, and so an iterative method was developed for obtaining the required initial parameters
for single-spin precessing systems. The initial data generation method is described in detail Appendix.~\ref{sec:ini_dat_gen}. Sec.~\ref{sec:sim_details_2} gives the details of the simulations with the parameters of all the simulations described in Tab:~\ref{Tab:nr_waveforms_list_ch4}.

\subsection{Details of the Simulations}~\label{sec:sim_details_2}
The simulations are split into three sets based on the mass-ratio of the system: $q=2$, $q=4$ and a super-kick series at $q=1$. 

The $q=2$ series is a set of four $q=2$ NR waveforms with a total in-plane spin of (dimensionless) magnitude 
$\chi_2 = S_2/m_2^2 = 0.7$, with $\theta_{SL} = \pi/2$ and $\phi_{\rm{Sn}}$ = (0, $\pi/2$, $\pi$, $3\pi/2$). 
For the $q = 4$ series, the spin is $\chi_2 = 0.8$, with the same $(\theta_{SL}, \phi_{\rm{Sn}})$ configurations as for the $q=2$ series. 

The $q=1$ simulations are two-spin systems in the ``super-kick'' 
configuration, where both black holes are spinning, with equal and 
opposite in-plane spins of $\chi_{i} = 0.8$. The super-kick configurations are non-precessing and due to the symmetry of the system, 
the final recoil is along $\pm \hat{z}$. 

For the simulation names, the following convention is used: q(mass-ratio of system)a(total spin of system)p(value of $\phi_{\rm{Sn}}$), 
following which, the first simulation in the $q=2$ series is q2a07p0. The angle, $\theta_{SL}$, between $\hat{L}$ and $\hat{S}$ is 
always $\pi/2$ for these systems, i.e., $\mathbf{S}_i^{\parallel} = 0$. For the $q=1$ series waveforms, remember that the total 
spin satisfies $\mathbf{S}_{1} + \mathbf{S}_{2} =0$, but we follow the above naming convention with "sk" in subscript for simplicity. 
We also use three 
extra NR simulations with different total in-plane spin magnitudes (with same $\theta_{SL}$ and $\phi_{\rm{Sn}} =0$), which are used as 
comparison cases, and were produced as part of the waveform catalogue presented in Ref.~\cite{bam_catalogue_paper}.

For the $q=2$ series, once the parameters for the $\phi_{\rm{Sn}}=0$ configuration were obtained, the parameters for the other 
simulations in the series were obtained simply by rotating the initial spin in the plane; the resulting eccentricities were all within our
tolerance. For the $q=4$ and $q=1$ series, however, the initial-parameter code was run separately for each value of $\phi_{\rm{Sn}}$. 

Initial momenta consistent with low-eccentricity inspiral were estimated using the PN/EOB evolution code described in 
Refs.~\cite{Hannam:2010ec,Purrer:2012wy,Husa:2015iqa}, with modifications as discussed in Appendix.~\ref{sec:ini_dat_gen}. 
We perform a short simulation of less than $1000M$ duration, and 
estimate the eccentricity from the co-ordinate separation, as given in Eq.~(3) of Ref.~\cite{PhysRevD.77.044037}.
For the $q=2$ and $q=1$ series, the eccentricities were all $< 5 \times 10^{-3}$, and we used the same initial momenta for 
production simulations. For the $q=4$ configurations, however, further eccentricity reduction was required. 

Ref.~\cite{Purrer:2012wy} describes an efficient procedure to further reduce eccentricity for non-precessing binaries. 
For the precessing simulations used here, we adopted a simpler procedure:
we performed a series of simulations with momenta increased or decreased by multiples of 0.1\%, until an eccentricity below our 
threshold was obtained. 
Note that the eccentricity for a system with $\phi_{\rm{Sn}} \rightarrow \phi_{\rm{Sn}} \pm \pi$ has the same value. 

BAM's mesh-refinement scheme is constructed as described in Refs.~\cite{PhysRevD.77.024027,Husa_2008}. In particular, a nested 
set of boxes centred on each black hole. For each simulation in this series, the boxes around the BHs consisted of 80 points 
in each direction, with a grid-spacing on the finest level of $m_1/56$, $m_1/36$ and $m_1/44$ for the $q=1$, $q=2$ and
$q=4$ series respectively. 
Further details of the grid setups are provided in Ref.~\cite{bam_catalogue_paper}. 
For two of the cases (q2a07p0 and q2a07p90), we performed higher-resolution runs 
with 96-point boxes, and a finest-level resolution of $m_1/48$. 
Using these two waveforms, we computed the 
match between the different resolution runs over a range of $(\theta, \phi)$ values (see Eq.~(\ref{eq:strain_decomposed})) using only 
the $l=2$ modes (as these are the modes used throughout the paper). We find that over the range of $(\theta, \phi)$ values considered, 
we obtain matches of $\sim$ 0.9995 - 0.99995. This shows that using the 96 point runs will not qualitatively change our results, but we 
will discuss this in more detail in Sec.~\ref{sec:setup}; see discussion pertaining to Fig.~\ref{fig:s1c_s2c_s3c}.

\section{Analysis methods}~\label{sec:methods}
This section provides the details of the match computation procedure employed for computing matches between the various 
precessing waveforms and the connection between the match and the SNR at which the template and signal can be distinguished
from each other. This is the primary method we use to interpret the results in Sec.~\ref{sec:results}.

\subsection{Match computations}~\label{sec:match_computation}

For the given physical system (with fixed intrinsic parameters), the detector response is uniquely determined by the system's
sky-position, inclination ($\iota$), coalescence-phase ($\phi_c$), polarisation ($\psi$) and time of arrival ($t_c$). The level 
of agreement between two gravitational waveforms can be ascertained by computing the match, $\mathcal{M}$, between the 
two waveforms. A value of $\mathcal{M}$ = 1 implies the waveforms are in perfect agreement. The smaller the value of 
$\mathcal{M}$, the larger the disagreement between the two waveforms. 

For a GW source directly overhead the detector, i.e., $(\alpha, \delta) = (0,0)$, the real valued detector response 
($h_{det}(t,\vec{\lambda})$), in terms of the two gravitational wave polarisations is,
	\begin{equation}~\label{eq:td_det_resp}
	h_{det}(t,\vec{\lambda}) = h_{+} \mathrm{cos}(2 \psi) + h_{\times} \mathrm{sin}(2 \psi) = \mathrm{Re}\left[ h(t,\vec{\lambda}) e^{2i \psi} \right],
	\end{equation} 
with $h_{+}$ and $h_{\times}$ as defined in Eq.~(\ref{eq:strain_decomposed}). Here, due to $(\alpha, \delta) = (0,0)$, the individual detector response depends only on $\psi$.

For precessing systems, the match between the signal ($\tilde{h}_{s}(f)$) and template  ($\tilde{h}_{t}$) waveform is given by~\cite{Schmidt:2014iyl},
	
		\begin{equation}~\label{eq:MaxMatch1}
	\begin{split}
	\underset{\sigma}{\max} \left<  \frac{\tilde{h_s}(f)}{||\tilde{h_s}(f)||} \middle\vert\ \frac{\tilde{h_t (f)}}{||\tilde{h_t (f)}||} \right> 	=  \frac{M}{||\tilde{h_s (f)}||} \sqrt{ \frac{N_1 - N_2 \mathrm{cos}(\sigma_n + 2 \sigma_m) }{N_{1}^2 - N_{2}^2}} .
	\end{split}
	\end{equation}
	where,
	\begin{equation}\label{eq:terms1}
N_{1} = \int_{-\infty}^{\infty} \frac{|\tilde{h_t}(f)|^{2}}{S_{n}(|f|)} \mathrm{d}f , \hspace{2mm}   N_{2} e^{i \sigma_{n}} =  \int_{-\infty}^{\infty} \frac{\tilde{h_t}(f) \tilde{h_t}(-f)}{S_{n}(|f|)} \mathrm{d}f .
\end{equation}
and 
	\begin{equation}~\label{eq:match_final}
	M e^{i \sigma_{m}} =  \int_{-\infty}^{\infty} \frac{\tilde{h}^{*}_{t} (f)}{S_{n}(|f|)}  \left[ \tilde{h}_{s}(f) e^{2i \psi} + \tilde{h}^{*}_{s}(-f) e^{- 2i \psi} \right] ,
	\end{equation}
	
	Given a template and signal waveform with given signal polarisation, Eq.~(\ref{eq:MaxMatch1})-Eq.~(\ref{eq:match_final}) give the match optimized 
over template polarisation. For further details on computation of match as above, see Appendix B of Ref.~\cite{Schmidt:2014iyl}.

For the match computations in this study, the $\phi_{\rm{Sn}} = 0$ system from each of the $q = 1$, $q = 2$ and $q = 4$ series of NR simulations
is used as the \emph{proxy} template with the other waveforms in each corresponding series as the signal waveforms. For the match computations, 
the signal is recomposed from 
only the $\ell=2$ modes using Eq.~(\ref{eq:strain_decomposed}) and Eq.~(\ref{eq:td_det_resp}). Each signal is uniquely defined by its 
inclination $\theta^{s}$, phase $\phi^{s}$, and polarisation $\psi^{s}$. (Note that we are considering a single detector network with the 
sky-position of the system exactly overhead the detector. Hence the angles $(\theta, \phi)$ can be interpreted as the inclination and phase 
with respect to the detector.)
For each unique signal, the match is maximised over the 
template $(\theta^{t}, \phi^{t}, \psi^{t})$. A total mass of $100 M_{\odot}$ is used for both signals and templates, PSD used is the \texttt{aLIGOZeroDetHighPower} PSD from \texttt{LALSimulation} package of \texttt{LALSuite} and the match is computed with $(f_{min} , f_{max}) \in $ (20, 600) Hz. 

For each system, the signals ($\theta^{s}, \phi^{s}$) are isotropically distributed over a sphere with 30 points in 
$\theta^{s}$ and 25 points in $\phi^{s}$. For each signal ($\theta^{s}, \phi^{s}$), we choose four values of $\psi^{s} \in [0, \pi/2)$ and 
then maximise the match over the template  ($\theta^{t}, \phi^{t}, \psi^{t}$). The match maximisation procedure goes through the 
following four steps,
	\begin{itemize}
	\item Isotropically grid the template $(\theta^t, \phi^t)$ space over the sphere with 41 points in $\theta^t$ and 81 points in $\phi^t$. 
	\item For each value of template $\theta^{t}_{i}$,  we compute the match across template $\phi^{t}_{j}$. For each ($\theta^{t}_{i}, \phi^{t}_{j}$) combination, the code gives the match optimized over template $\psi^{t}$. 
	\item For each $\theta^{t}_{i}$, the match is interpolated over the $\phi^{t}_{j}$ values, from which the maximum match over $\phi^{t}$ for each $\theta^{t}_{i}$ is obtained. 
	\item Thus, we get a set of match values across the template $\theta^{t}_{i}$ values, which are then interpolated to obtain the maximum match over template $(\theta^t, \phi^t, \psi^t)$. 
	\end{itemize}   	
The choice of 41$\times$81 grid for the template waveforms for match maximisation was chosen by balancing the i) accuracy of 
final result and ii) computational time required for each match computation. Using a few random signal $(\theta, \phi)$ values, we found 
that doubling the grid size changed the results by at most $5\%$ while doubling the computational cost. 


\subsection{Confidence intervals from match values}~\label{sec:match_connection}

Given two waveforms close to each other in the parameter space, i.e., $h_{1}(\vec{\lambda_1})$ and $h_{2}(\vec{\lambda_2})$ such that $\vec{\lambda_1} \sim \vec{\lambda_2}$, where we have the SNR ($\rho$) of the signals and the match ($\mathcal{M})$ between the two, both waveforms will have consistent posterior distributions within 90\% confidence interval of each other if,  
	\begin{equation}
	\mathcal{M}[h_{1}, h_{2}] \geq 1 - \frac{\chi_{k}^2 (1-p)}{2 \rho^{2}}.
	\end{equation}
	Alternatively, two waveforms would be distinguishable from each other if the posteriors recovered for the two have different confidence intervals. So, given a match value, the above gives us a condition for the SNR ($\rho_c$) at which the waveforms would be distinguishable, 
	\begin{equation}~\label{eq:det_snr}
	\rho_c \geq \sqrt{ \frac{\chi_{k}^2 (1-p)}{ 2(1 - \mathcal{M}) } } .
	\end{equation}
	For the systems under consideration, there are a total of 7 parameters which can be varied and hence, $k=7$. At $k=7$ for 90\% confidence intervals, $\chi_{k}^2 (1-p) = 12.02$.

\begin{figure*}
\includegraphics[width=0.95\textwidth,keepaspectratio]{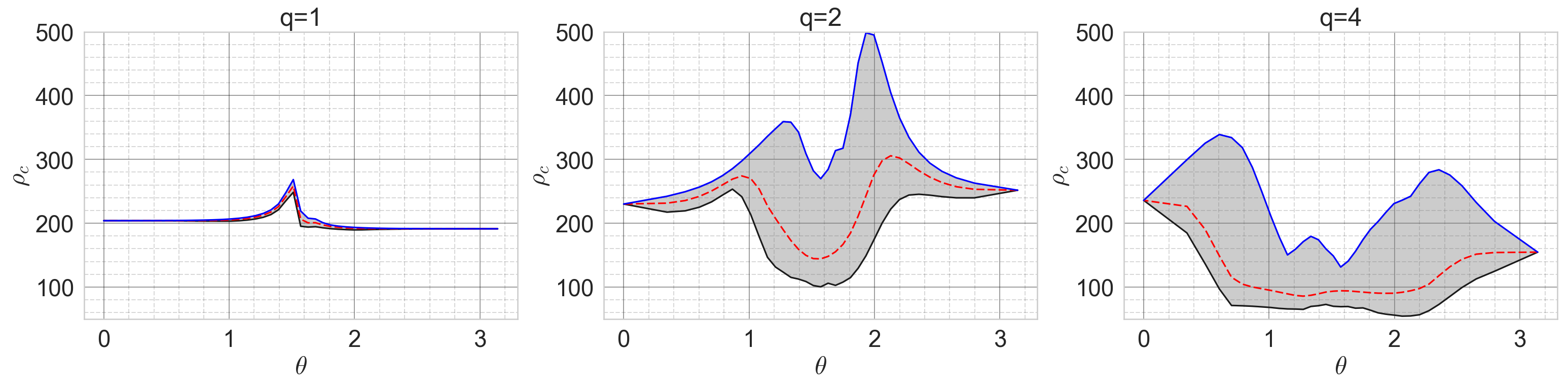}
\caption{The left, middle and right columns show $\rho_c$ for signals q1a08p180$_{\rm{sk}}$, q2a07p180, and q4a08p180 as seen by templates q1a08p0$_{\rm{sk}}$, q2a07p0 and q4a08p0 respectively, across the signal $\theta$ space. For each signal $\theta$, the match is computed with the template at $\theta + \pi$, over a range of $\phi$ values, and the 
black, dashed-red and blue lines show the minimum, mean and maximum match ($\rho_c$) across the $\phi$ space. 
We observe a larger variation of $\rho_c$ for the $q=2$ and $q=4$ cases as compared to the $q=1$ due to presence of non-zero 
subdominant modes, ($l,m$) = (2,1), (2,0) and (2,-1).  
} ~\label{fig:s1c_s2c_s3c}
\end{figure*}

See Ref.~\cite{Baird:2012cu} for a detailed discussion for the condition used above, although previous 
studies~\cite{Flanagan:1997kp, Lindblom:2008cm, McWilliams:2010eq, Cho:2012ed, Chatziioannou:2017tdw} have used similar definitions to 
determine the distinguishability/accuracy requirements of gravitational waveforms. As was pointed out in Ref.~\cite{Purrer:2019jcp}, the equality 
in Eq.~(\ref{eq:det_snr}) is a sufficient, but not always a necessary condition, to determine the accuracy between two waveforms. For a given signal 
and template waveform with maximum match $\mathcal{M}$ and corresponding $\rho_c$, if the signal $\rho < \rho_c$, biases due to detector noise 
will dominate biases due to model systematics.  Alternatively, if the opposite is true, biases \emph{may} arise during parameter inference and so, 
the above equality is a conservative estimate of accuracy requirements. Hence, using the set of match values computed from 
Sec.~\ref{sec:match_computation} for each system and Eq.~(\ref{eq:det_snr}), we can then estimate the SNR $(\rho_c)$ at which the signal 
system could be distinguished by the proxy template.


\section{Organisation of results}~\label{sec:setup}

In subsequent sections we study the SNR ($\rho_c$) at which configurations with differing spin directions or spin magnitudes 
are distinguishable, as defined by Eq.~(\ref{eq:det_snr}). 
In this section we make some general comments on the accuracy of $\rho_c$ for our simulations, some general properties of the waveforms
with respect to changes in the in-plane spin direction, $\phi_{\rm{Sn}}$, and examine how $\rho_c$ varies with respect to different 
binary orientations. This motivates the way we will present our results for the remainder of this paper. 

Let us first discuss accuracy. As reported in Sec.~\ref{sec:simulation_details}, the matches between the 80- and 96-point runs 
(for q2a07p0 and q2a07p90)  
are $\sim$0.9995 - 0.99995, which translates to $\rho_c$ between 110 and 350. This suggests that we can identify two waveforms
as indistinguishable up to SNRs of at least 110. We also computed the matches between the $\phi_{\rm{Sn}} =0, \pi/2$ 
systems using the corresponding 80- and 96-point waveforms over a range of $(\theta, \phi)$ values, and found that the relative error between 
them is $\mathcal{O}(0.05\%)$. These numbers suggest that although we should be cautious when interpreting 
very large values of $\rho_c$, we expect the qualitative behaviour of the matches to remain unchanged with more 
accurate simulations. 
	
Given the accuracy limits of our simulations, we are now in a position to study how the waveforms vary with respect to different initial
directions of the in-plane spin. We begin by noting an approximate symmetry between systems with a $\phi_{\rm{Sn}}$ difference 
of $\pi$. An in-plane spin rotation of $\pi$ corresponds to flipping the direction of the out-of-plane recoil, and therefore we
would expect that the signal from a system with a given value of $\phi_{\rm{Sn}}$ to be identical to that from a system 
with $\phi_{\rm{Sn}} + \pi$, if observed from the opposite side of the orbital plane, i.e., with $\theta \rightarrow \theta + \pi$. 
We have verified that the optimal match is 
indeed found when $\theta_{\rm{template}} \approx (\theta_{\rm{signal}} + \pi$). In 
 Fig.~\ref{fig:s1c_s2c_s3c} we plot the $\rho_c$ across signal $\theta$ for a range of signal $\phi$ for which $\mathcal{M} $ 
 is computed with $(\theta_{\rm{template}},\phi_{\rm{template}}) = (\theta_{\rm{signal}} + \pi, \phi_{\rm{signal}})$ and the match is 
 optimized only over template $\psi$. The $\rho_c \geq 100$ for all $\phi_{\rm{Sn}} = \pi$ signals for the $q=1$ and $q=2$ 
 systems, with the $\rho_c \geq 50$ for each  $\theta_{\rm{signal}}$ for $q=4$ system. 
We do not observe an exact symmetry (i.e., a mismatch of zero), because in the $q=1$ and $q=4$ systems, we did not simply rotate 
 the spin between each configuration, but instead calculated initial parameters individually for each value of $\phi_{\rm{Sn}}$, 
 so these do not form a one-parameter family. Even with rotated spins within a series, as in the $q=2$ series, we have  not changed the momenta; we would expect even lower mismatches if the out-of-plane momenta, in the initial data, had been reflected in the orbital plane between the $\phi_{\rm{Sn}}$ and $\phi_{\rm{Sn}}+\pi$ configurations. 

 \begin{figure}
	\includegraphics[width=0.48\textwidth,keepaspectratio]{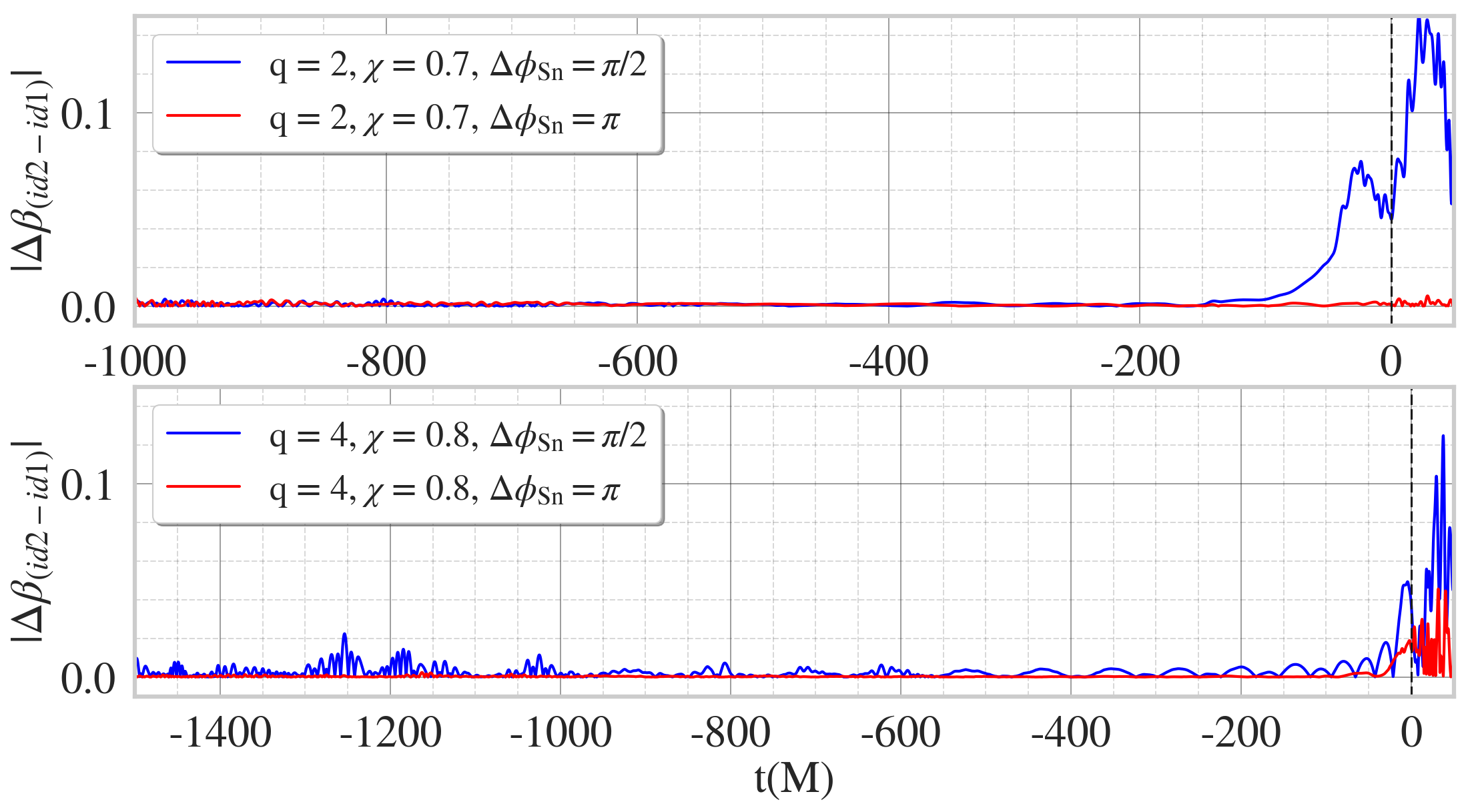}
	\caption{ Top Panel : Difference in the $\beta$ Euler-angle (in radians) between the $\phi_{\rm{Sn}}=0, \pi/2$ (Blue) and $\phi_{\rm{Sn}}=0, \pi$ (Red)  configurations of the $q=2, \chi=0.7$ system. Bottom Panel : The same as above, but for the $q=4, \chi=0.8$ system. The legend gives the mass-ratio and spin of signal waveform and the parameter varied between the signal and template waveform. For both systems, $\Delta \beta$ is small during late-inspiral, with the majority of differences arising near merger.
	}~\label{fig:beta_angle_differences}
	\end{figure}	
	
 \begin{figure*}
	\includegraphics[width=0.48\textwidth,keepaspectratio]{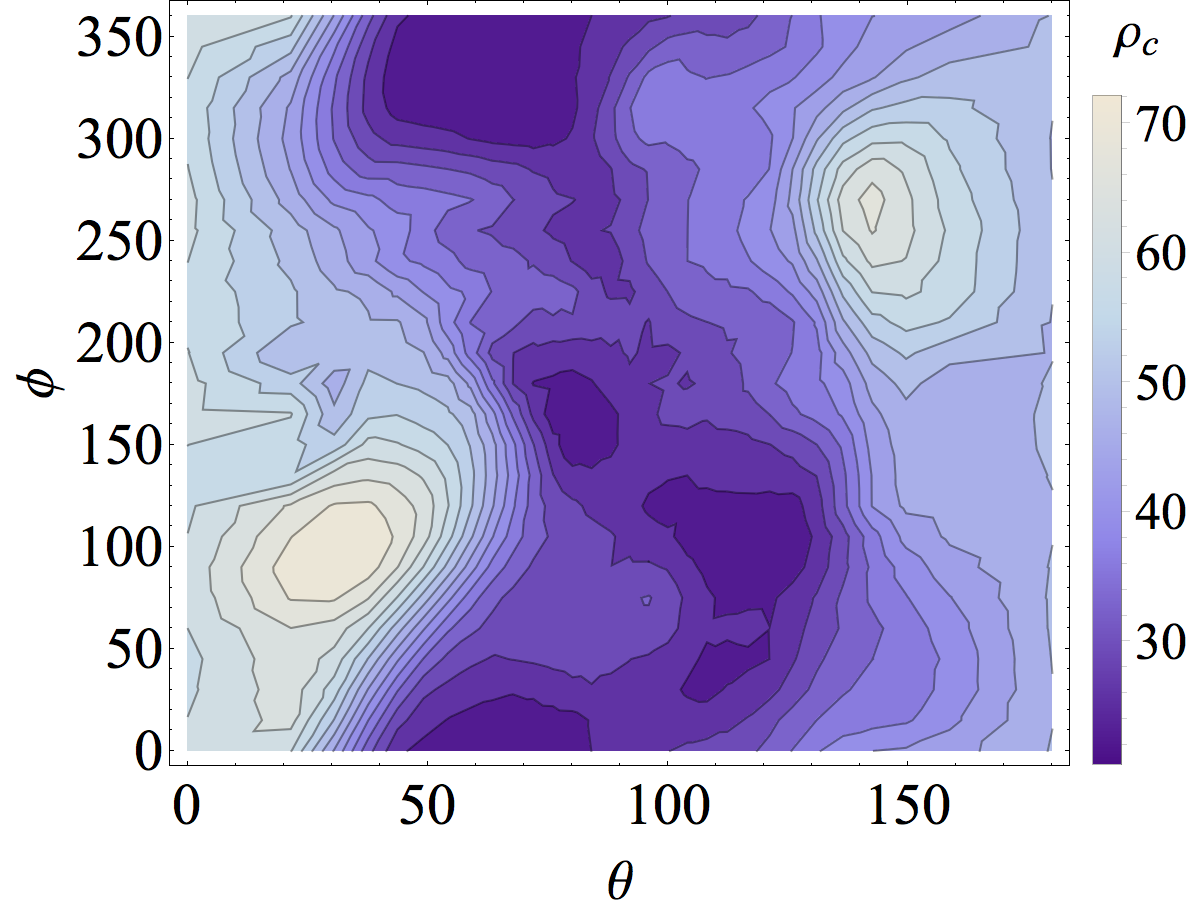}
	\includegraphics[width=0.48\textwidth,keepaspectratio]{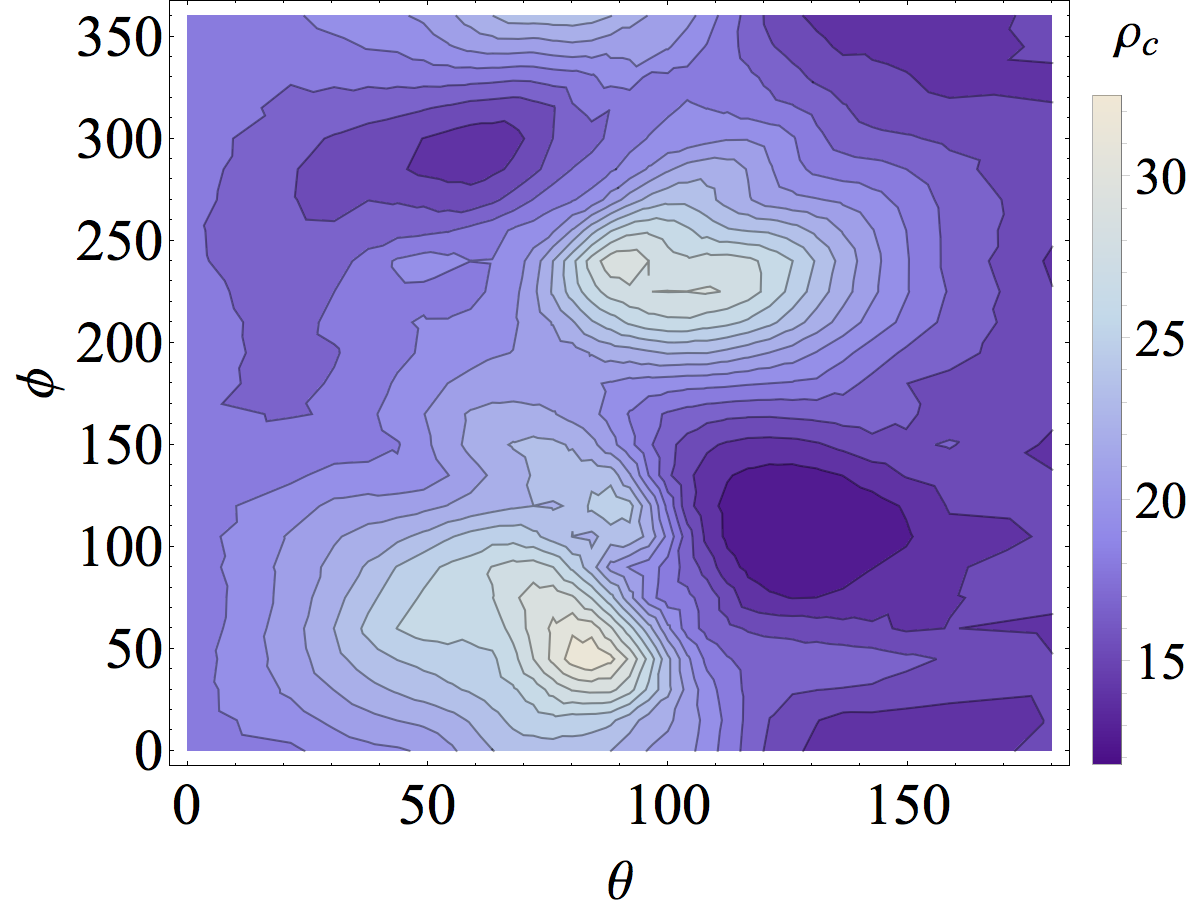}
	\caption{The left and right panels show the contour plot of $\rho_c$ for the signal q2a07p90 and q4a08p90 as seen by template q2a07p0 and q4a08p0 respectively over the signal $(\theta, \phi)$ values with the label on the right of each plot showing the values of $\rho_c$. See text for further discussion. }~\label{fig:2d_contour_q2q4}
	\end{figure*}
	
When we rotate the initial in-plane spin direction $\phi_{\rm{Sn}}$, this will, by definition, rotate the initial orbital angular momentum about 
the total angular momentum, i.e., a change in the $\alpha$ Euler angle in Eq.~(\ref{eq:wigner_rotation}). But we do not expect it to have 
a significant impact on the opening angle $\beta$ between the orbital and 
total angular momenta, at least during the inspiral. In Fig.~\ref{fig:beta_angle_differences} we plot the differences in the $\beta$ Euler 
angle for the $q=2$ and $q=4$ systems with differences in $\phi_{\rm{Sn}}$ of $\pi/2$ and $\pi$. 
We can see that $\Delta \beta \sim 0$ for $q=2$, but rises to a few $\sim \mathcal{O}(1^{\circ})$ for $q=4$; this is consistent with our 
expectation as the construction of the $q=2$ series configurations better represent the symmetry. We note that in 
comparing the $\phi_{\rm{Sn}} = 0$ and $\phi_{\rm{Sn}} = \pi/2$ configurations, there is a clear difference in $\beta$ during the
merger and ringdown. 
	
 We now wish to estimate how easily two configurations can be distinguished for different choices of binary orientation, 
$(\theta,\phi)$. In all of these comparisons, we average $\rho_c$ over four choices of waveform polarisation. 
As mentioned in Sec.~\ref{sec:match_computation}, for each signal system we have ($25 \times 30 \times 4 = 3000$) match values and the 
SNR for each. To average across different choices of signal polarisation, $\psi^s$, we follow 
Refs.~\cite{Khan:2018fmp,Harry:2017weg, Buonanno:2002fy} and average the match for each $(\theta^s , \phi^s)$ across the 
signal polarisation $\psi^s$ by weighting them with their SNR. This approximately accounts for the likelihood of the signal being 
detected. This SNR-averaged match is defined as, 
	\begin{equation}~\label{eq:snr_aveg_match}
	\overline{\mathcal{M}} = \left(  \frac{\sum_{i} \rho_{i}^{3} \mathcal{M}_{i}^{3}}{\sum_{i} \rho_{i}^{3}} \right)^{1/3},
	\end{equation}
where the sum is over all four signal polarisation values. 
So, for a given system, we have 750 values of the SNR averaged match.

Fig.~\ref{fig:2d_contour_q2q4} shows contour plots of the variation of $\rho_c$ across the signal $(\theta, \phi)$ for the q2a07p90 
and q4a08p90 signals as seen by the q2a07p0 and q4a08p0 systems respectively, where the match at each $(\theta, \phi)$ point is 
maximised over template $(\theta, \phi, \psi)$ and 
then averaged over the signal $\psi$ values using Eq.~(\ref{eq:snr_aveg_match}) and $\rho_c$ is computed using Eq.~(\ref{eq:det_snr}). 
For these systems with $\phi_{\rm{Sn}}$ differences of $\pi/2$, the major contribution to the mismatch would be from a combination of their 
slightly different precession motion (as seen from Fig.~\ref{fig:beta_angle_differences}) and mode-asymmetry behaviour. For the $q=2$ 
system, $20 \lesssim \rho_c \lesssim 72$, whereas for the $q=4$ system,  $11 \lesssim \rho_c \lesssim 32$. Precession effects are more 
pronounced at edge-on than face-on inclination, so the lower $\rho_c$ for q2a07p90 at $\theta \sim \pi/2$ is expected. For q4a08p90, 
this behaviour seems to reverse (higher $\rho_c$ at edge-on compared to face-on). In Sec.~\ref{sec:inertial_symm_wvfm_results}, 
we remove the mode-asymmetry from the waveforms and compute the matches between the symmetrized waveforms for the 
$\phi_{\rm{Sn}} = \pi/2$ signals as seen by $\phi_{\rm{Sn}} = 0$ template. For the $q=2$ system, the band of low $\rho_c$ near edge-on inclination broadens for all $\phi$ and for the $q=4$ system, the peaks of high $\rho_c$ shift slightly away from near edge-on inclinations with an overall increase in $\rho_c$ for both cases. This implies that the behaviour of $\rho_c$ across the ($\theta, \phi$) space can be strongly affected by mode-asymmetric content for these systems.

We see from Fig.~\ref{fig:2d_contour_q2q4} that there is a wide variation in the SNR at which these configurations would be 
distinguishable, depending on their orientation to the detector. Also, the exact way in which precessional motion and mode-asymmetry 
affect distinguishability over the $(\theta, \phi)$ space is hard to characterize. For the main results in this paper, we find it more instructive to 
use a measure that would give an idea of the variation of $\rho_c$ across all orientations. For a given signal and template configuration, we 
define the quantity $\Gamma (\rho)$ which gives the percentage of signals distinguishable at a given SNR by the template. This quantity can 
be defined formally as,
\begin{equation}~\label{eq:cumulative_dist_func}
	\Gamma (\rho) = 100 \frac{\text{len}( \mathcal{S}_{\text{id2} : \text{id1}} [\rho_c < \rho])}{\text{len}( \mathcal{S}_{\text{id2} : \text{id1}})} ,
	\end{equation}        
where $\mathcal{S}_{\text{id2} : \text{id1}} [\rho_c < \rho]$ is the set of all signals with distinguishability SNR ($\rho_c$) smaller 
than a given SNR value ($\rho$) and $\mathcal{S}_{\text{id2} : \text{id1}}$ is the set of all the available signals. In the following sections, this 
cumulative measure of the fraction of signal distinguishable at a given SNR or lower, will be used as our main tool to quantify the 
differences between binary configurations.


\section{Results}~\label{sec:results}

We now consider in detail the distinguishability of our NR configurations. To reiterate, for each system at a given mass-ratio, the $\phi_{\rm{Sn}} = 0$ system is used as the proxy template waveform. For the results hence, compared to the proxy templates, the signals either have a different spin direction or the magnitude. Also, due to the approximate symmetry between the $\phi_{\rm{Sn}} \pm \pi$ systems, the results for the $\phi_{\rm{Sn}} = 3\pi/2$ and $\phi_{\rm{Sn}} = \pi/2$ are very similar, and hence, we will only present results for the $\phi_{\rm{Sn}} = \pi/2$ systems. 


	\begin{table}[t]
	\begin{tabular}{|c|c|c|c|c|c|}
	\hline 
	Template & Signal & Average Match & $\rho_c$  \\
	\hline
	$q=2,\chi=0.7,\phi_{\rm{Sn}}=0$ & $\phi_{\rm{Sn}} +\pi/2$ & 0.9983 & 60 \\
	``q2a07p0'' & $\phi_{\rm{Sn}} + \pi$  & 0.9999& 250 \\
	
	 & $\chi + 0.1$  & 0.9983 & 60 \\
	 & $\chi - 0.3$  & 0.9952 & 36 \\
	\hline
	\hline
	$q=4,\chi=0.8,\phi_{\rm{Sn}}=0$ & $\phi_{\rm{Sn}} + \pi/2$  & 0.9811 & 17 \\
	``q4a08p0'' & $\phi_{\rm{Sn}} + \pi$  & 0.9997 & 143 \\
	 & $\chi - 0.4$  & 0.9936 & 30 \\
	\hline
	\hline
	$q=1,\chi=0.8,\phi_{\rm{Sn}}=0$ & $\phi_{\rm{Sn}} + \pi/2$   & 0.9981 & 57 \\
	``q1a08p0$_{\rm {sk}}$'' & $\phi_{\rm{Sn}} + \pi$   & 0.9998 & 197 \\
	\hline
	\end{tabular}
	\caption{Match and distinguishability SNR $\rho_c$ between different configurations, averaged over all $(\theta, \phi, \psi)$ values.
	We consider three template waveforms (left), and a variety of different signals. See text for further discussion.
	}~\label{Tab:system_match_full_t1}
\end{table}

We first summarise the main differences that we observe between the different configurations, by averaging the match and distinguishability 
SNR $\rho_c$ over all 3000 orientations and polarisations. The results are shown in Tab.~\ref{Tab:system_match_full_t1}. 
We see that variations in the in-plane spin direction can be distinguishable at an SNR of 60, and sometimes as low as $\sim$20. 
Similarly, waveforms from systems with different spin \emph{magnitudes} can be distinguishable are SNRs of $\sim$60 for spin 
differences on the order of 0.1. These results encapsulate the two key conclusions we derive from this study: differences in the 
waveforms between configurations with different in-plane spin directions may be measurable with observations in the near future
(the highest BBH SNR to date has been GW150914, with an SNR of $\sim$24~\cite{TheLIGOScientific:2016pea}), and the SNRs
at which in-plane spin magnitudes could be measured are comparable to those at which the spin direction will also impact the results. 
This strongly suggests that waveform changes due to the in-plane spin direction (beyond an overall offset in the precession angle $\alpha$)
need to be included in waveform models. 

These remainder of this paper considers these results in more detail, and we also attempt to isolate the physical effects that lead
to these waveform differences. 


In Sec~\ref{sec:full_wvfm_results} we compare the full NR waveforms (using all the $\ell=2$ multipoles). This allows us to identify the 
range of SNRs in which the configurations with different choices of $\phi_{\rm{Sn}}$ will be distinguishable, and to compare this with the 
effect of changing the  in-plane spin magnitude. 

We then attempt to isolate the causes of these differences. In Sec.~\ref{sec:qa_symm_wvfm_results}, we transform the waveforms into 
the co-precessing frame (where modes with $|m|<2 \approx 0$) and study the matches between the waveform with symmetrized 
$(l, |m|) = (2,2)$ modes for $\phi_{\rm{Sn}} \pm \pi/2$ systems. This allows us to estimate the distinguishability of two waveforms when 
both precession and mode-asymmetry effects are muted, due primarily to small differences in the inspiral rate and merger-ringdown 
differences. These co-precessing-frame symmetrized modes are then transformed back to the inertial frame and
Sec.~\ref{sec:inertial_symm_wvfm_results} presents the results of analysis with those waveforms. These results estimate the impact of 
neglecting mode-asymmetry on the distinguishability of precessing-binary waveforms.

\subsection{Full waveform analysis}~\label{sec:full_wvfm_results}

\begin{figure}
	\includegraphics[width=0.48\textwidth,keepaspectratio]{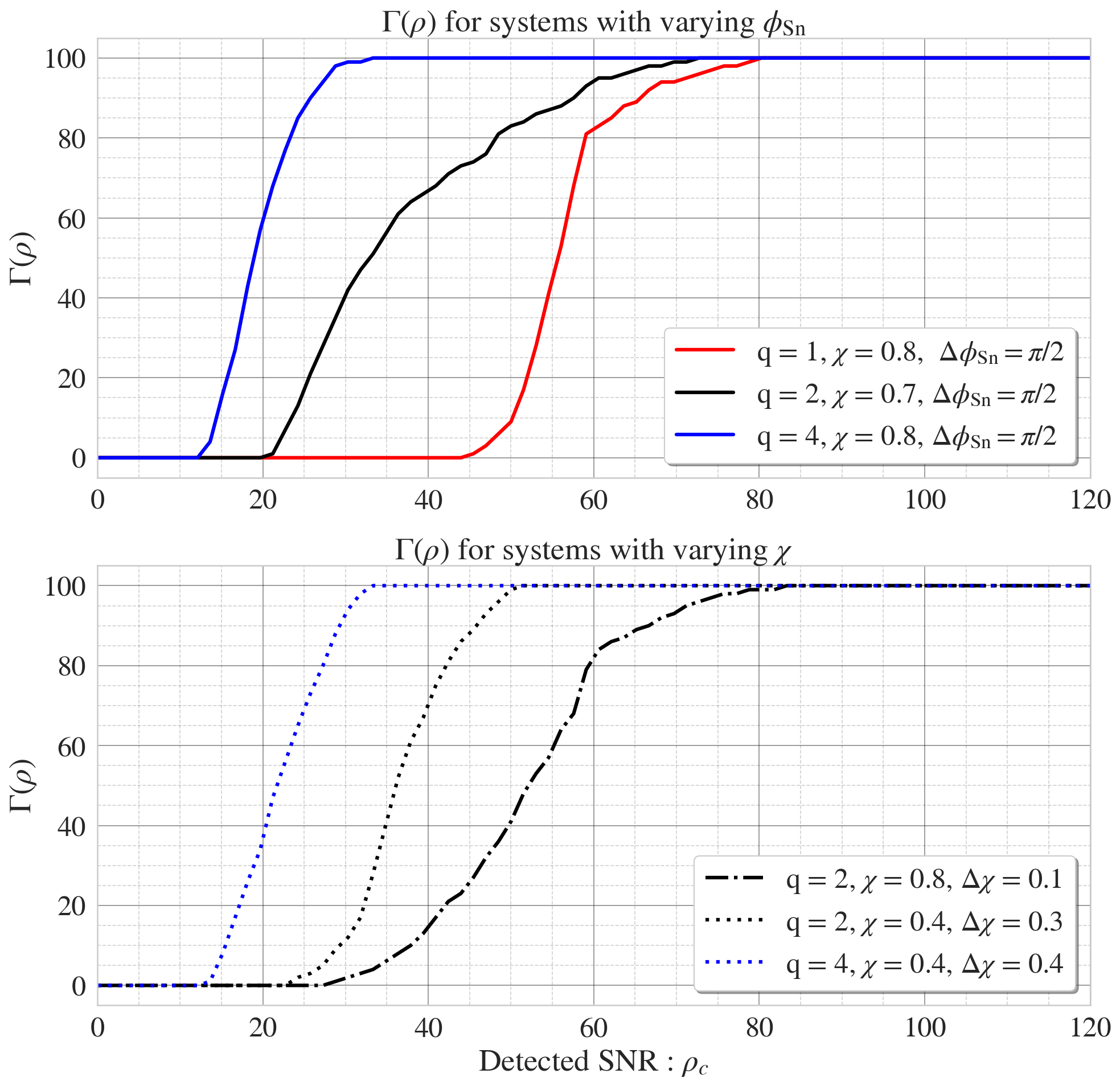}
	\caption{$\Gamma (\rho)$ as defined in Eq.~(\ref{eq:cumulative_dist_func}). Top panel: $q = 1,2,4$ systems with
$\phi_{\rm{Sn}}$ differences of $\pi/2$. Lower panel: $q = 2,4$ systems with different spin values. The results for $q=1$, $q=2$ and 
$q=4$ are shown in Red, Black and Blue respectively. 
The solid lines show the results for systems with $\phi_{\rm{Sn}}$ differences of $\pi/2$ and the same $\chi_p$, 
the dashed-lines show the results for large $\chi_{p}$ differences (0.3 for $q=2$ and 0.4 for $q=4$) with the same $\phi_{\rm{Sn}}$,
and the dotted-dashed for small $\chi_p$ differences (0.1 for $q=2$) with the same $\phi_{\rm{Sn}}$. See text for further details. 
}~\label{fig:s123b_fullvfull}
	\end{figure}

The key results of this work are shown in Fig.~\ref{fig:s123b_fullvfull}. The figure shows the percentage of signals with different spin direction (top panel) or magnitude (bottom panel) that will be distinguishable below a given SNR with the corresponding $\phi_{\rm{Sn}} = 0$ template for the  $q = 1, 2, 4$ systems. In the legend, we mention the mass-ratio and spin of the signal waveform and the parameter varied between the signal and template waveform. Unless mentioned otherwise, for all plots hence, systems with $q=1, 2, 4$ are colour coded with Red, Black and Blue respectively. 

For the $q=1$ super-kick configurations, the detectability between $\phi_{\rm{Sn}} \pm \pi/2$ systems is due to asymmetric radiation 
of gravitational modes. The detectable SNRs for these super-kick systems, $45 \lesssim \rho_c \lesssim 80$, are 
in the possible range of ground based detectors, but will be rare; we expect less than one in every hundred signals to have such high 
SNRs. The recoil velocities for the $q=1$ waveforms used here are $\sim$ 700 km/s 
($\phi_{\rm{Sn}}=0$) and $\sim$ 2700 km/s ($\phi_{\rm{Sn}}=\pi/2$). For systems with lower spins (and hence lower recoil velocities), 
we can expect larger values of $\rho_c$, meaning that these differences will be more difficult to measure. These results are 
consistent with those presented in Refs.~\cite{Varma:2020nbm, Lousto:2019lyf}.   

The $\rho_c$ for the $q=2$ systems with $\phi_{\rm{Sn}} = \pi/2$ are in the range of $20 \lesssim \rho_c \lesssim 75$,
and for $q=4$ they are $12 \lesssim \rho_c < 35$. Given that GW signals have already been observed with SNRs as high as 
30~\cite{TheLIGOScientific:2017qsa}, and the detection threshold is at an SNR of approximately 10, these are well within the range
of current ground-based detectors. We emphasize that these results do not mean we can necessarily measure, for example, the 
spin direction at the frequency when the signal enters the detector's sensitivity band; this quantity may be degenerate with other 
physical properties. However, they do indicate that systems with different values of $\phi_{\rm{Sn}}$ can be distinguished from each 
other, and if we do not take into account the effects on the waveform of varying $\phi_{\rm{Sn}}$ (as in current Phenom and EOB models),
then these differences will manifest themselves in biases in at least one physical parameter for sufficiently strong signals. 

We might expect that the effect of $\phi_{\rm{Sn}}$ on measurements will be far smaller than that of the spin magnitude. The lower panel
of Fig.~\ref{fig:s123b_fullvfull} shows that this is not necessarily the case. For example, the $q=2$ system with $\chi_p$ of 0.8, the distinguishability SNRs are in the range $30 \lesssim \rho_c \lesssim 80$. A change in the spin
of 0.1 is therefore, in general, slightly more difficult to distinguish than a change in the in-plane spin direction of $\pi/2$. A change in the 
spin magnitude of 0.3 (for the $q=2$, $\chi_{p} = $ 0.4 system) is distinguishable at SNRs in the range $25 \lesssim \rho_c \lesssim 45$.
As we can see by comparing with the upper panel, this is comparable to the distinguishability of a spin rotation of $\pi/2$. 
In the $q=4$ configurations, we see that a spin change of 0.4 (between 0.8 and 0.4) is distinguishable in the SNR range
$13 \lesssim \rho_c < 35$, again comparable to what we see for a spin rotation. These results suggest that the SNRs at which 
in-plane spin magnitudes become measurable are also the SNRs at which changes in the waveform due to spin rotations also 
become measurable. As noted above, this study cannot tell us which physical measurements will be biassed by models that 
neglect mode asymmetries or changes in the binary dynamics, but our results raise the possibility that accurate measurements of precessing 
systems, i.e., of black-hole spins, will not be possible without the inclusion of some or all of these effects in waveform models. 
	
For the $q=2$ and $q=4$ systems with different spin directions, we observe a slight difference in the merger times, mode-asymmetric 
content as well as precessional dynamics (as can be seen from the $\Delta \beta$ plot in Fig:~\ref{fig:beta_angle_differences}). 
These differences are the main reason for distinguishability of systems with different spin directions. These effects will become weaker
for lower spins, but one should bear in mind that precession effects and black-hole spins will also become more difficult to 
measure~\cite{Fairhurst:2019srr}. As such, we expect these results to be largely independent of spin magnitude. A more important caveat on these results
is that they are restricted to signals of total mass of 100\,$M_\odot$. For lower-mass systems we expect the mode asymmetries to 
contribute less to the SNR, and therefore to have less impact. We leave a study of the impact of mode asymmetries on parameter measurements to future work.

\subsection{QA frame symmetrized waveform analysis}~\label{sec:qa_symm_wvfm_results}

\begin{figure}
	\includegraphics[width=0.48\textwidth,keepaspectratio]{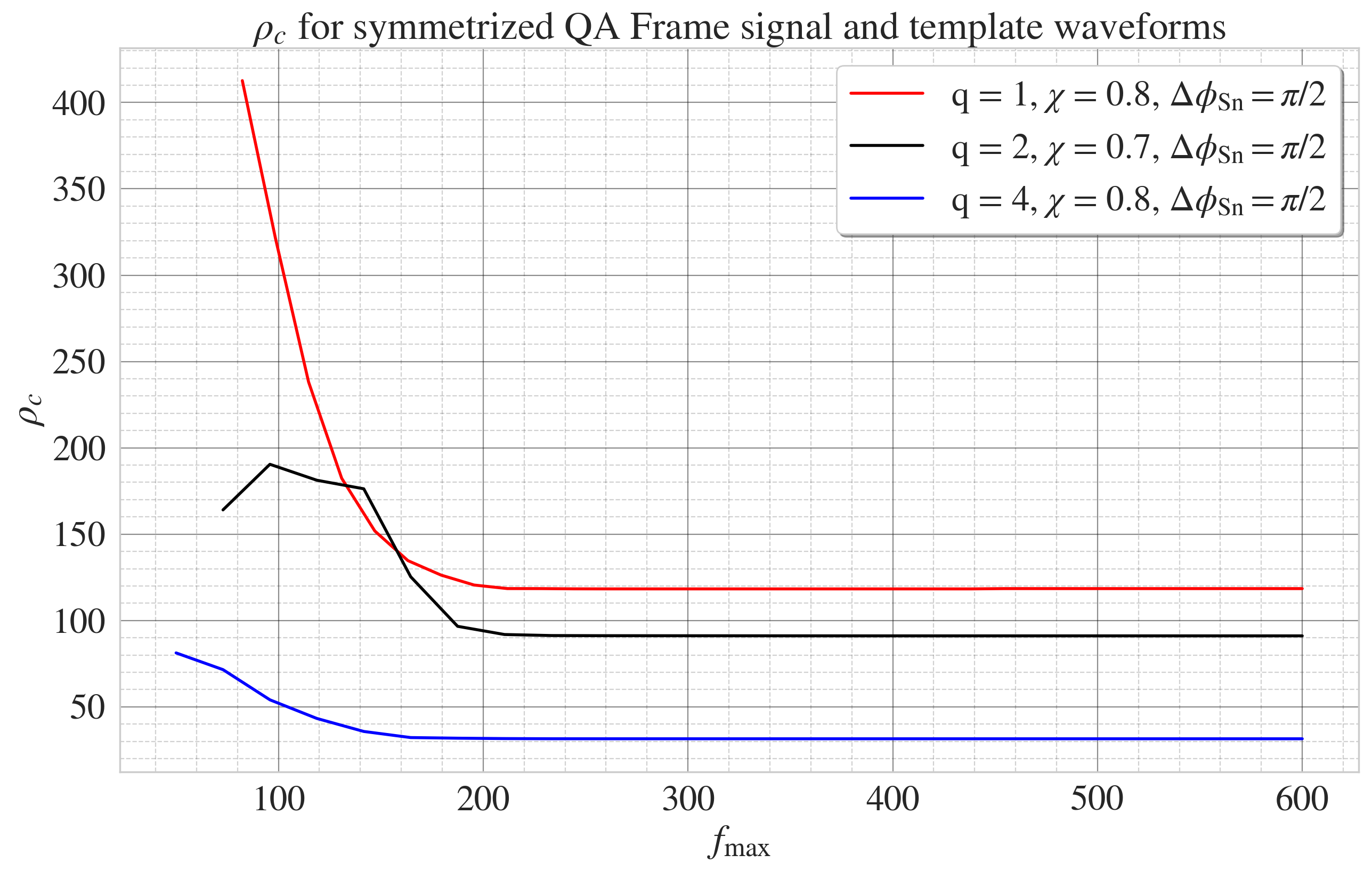}
	\caption{This plot shows the $\rho_c$ computed from the match ($\mathcal{M}$) between the symmetrized QA frame waveforms for the $q=2$, $q=4$ systems (solid-black, solid-blue respectively) and symmetrized $q=1$ waveforms (solid-red), with 
varying values of the upper cutoff frequency in $f_{max}$ for the match calculation. The legend follows the same naming convention as Fig:~\ref{fig:s123b_fullvfull}.  
	}~\label{fig:s123b_qa_symm}
	\end{figure} 
	
As mentioned previously, for the $q=2$ and $q=4$ systems with different $\phi_{\rm{Sn}}$, the mismatches are primarily due to 
differences in their precessional motion (i.e., differences in the 
$(\alpha, \beta, \epsilon)$ angles) and mode-asymmetric content. In this section our aim is to remove, as much as possible, the precession and mode-asymmetry effects, and to quantify the impact of all other effects (inspiral rate and merger-ringdown behaviour). We transform the $q=2$ and $q=4$ 
waveforms into the co-precessing frame (specifically, the quadrupole-aligned, ``QA'', frame~\cite{Schmidt:2012rh,Boyle:2011gg,OShaughnessy:2011pmr}) 
 using Eq.~(\ref{eq:wigner_rotation}). This minimises modulations due to precession. In 
this frame the dominant power is in the $(\ell=2,|m|=2)$ harmonics. We then symmetrise these harmonics, to remove the effects
of mode asymmetries. In terms of the QA frame modes $(h_{lm}^{QA})$, the symmetric waveform in the QA frame 
$(h_{22}^{QA,symm})$ is defined as, 
\begin{equation}~\label{eq:symm_wvfm_def}
h_{22}^{QA, symm} = \frac{1}{2} \left( h_{22}^{QA} + h_{2,-2}^{* QA} \right) ,
\end{equation} 
where $h^{*}_{l,m}$ is the complex conjugate of the mode. Using this, we can define the $(2,-2)$ mode as, 
$h_{2,-2}^{QA, symm} = h_{2,2}^{* QA, symm}$, using the relation $h_{\ell m} = (-1)^{\ell}h^{*}_{\ell,-m}$. 
Doing this for the $q=2$ and $q=4$ systems removes the precession modulations and mode-asymmetry. As the super-kick 
simulations are non-precessing, those waveforms are symmetrized in the inertial frame using Eq.~(\ref{eq:symm_wvfm_def}).

\begin{figure}
	\includegraphics[width=0.48\textwidth,keepaspectratio]{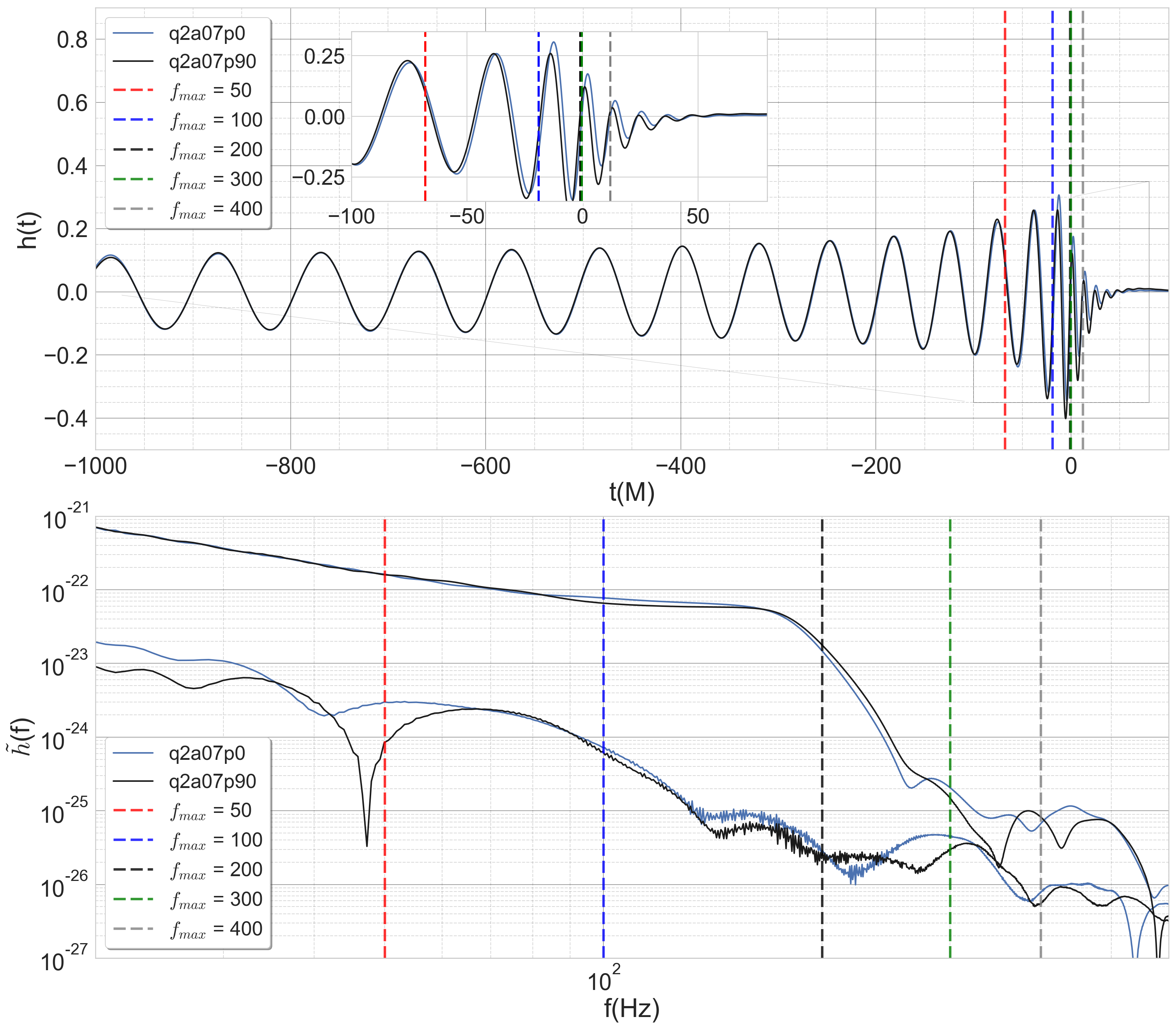}
	\caption{In the top [bottom] panel, we plot the q2a07p0 (blue) and q2a07p90 (black) time [frequency] domain QA frame symmetrized waveforms. For the top panel, the dashed lines show the time at which the waveform has a specific frequency used as $f_{max}$ value for Fig:~\ref{fig:s123b_qa_symm}. For the bottom panel, the dashed lines show the position of that frequency with respect to the frequency domain waveform. Frequency values of (50, 100, 200, 300, 400) are given in dashed (red, blue, black, green, gray) lines respectively.
	}~\label{fig:q2_series_td_fd_vs_fmax}
\end{figure}

Matches calculated between symmetrised QA $(2,2)$ modes are independent of orientation and polarisation, so the averaging that
we performed previously is no longer necessary. Between the $\phi_{\rm{Sn}} \pm \pi/2$ configurations at mass ratios
 $q = 1,2,4$, the indistinguishability SNRs are now 120, 90 and 30, respectively. If we contrast these with the top panel of 
 Fig.~\ref{fig:s123b_fullvfull}, we see that for the $q=2$ and $q=4$ cases, differences in the signal phase make a noticeable contribution to the 
 indistinguishability SNR. In Fig.~\ref{fig:s123b_qa_symm} we show $\rho_c$ over a range of $f_{max}$ values.
 Fig.~\ref{fig:q2_series_td_fd_vs_fmax} shows the $q=2$ series waveforms in time and frequency domain,
to illustrate where these choices of $f_{max}$ occur during the binaries' coalescence. These figures show that, as we might expect,
most of the disagreement between the waveforms accumulates during merger and ringdown.   

 These results should be taken with a few caveats. As already mentioned, for the $q=2$ waveforms obtained with 80- and 96-point 
 resolutions, over the $\theta$ space, the match lies between 0.9995 - 0.99995 which translates to $\rho_c$ of $\sim$ 110 - 345. 
 So, although the QA frame symmetrized matches are close to the minimum match due to NR uncertainties, over the majority of 
 the $\theta$ space, the QA frame symmetrized results should hold even for more accurate NR waveforms. For the $q=4$ system, 
 to obtain the low eccentricity parameters, the momenta between the $\phi_{\rm{Sn}}=0, \pi/2$ systems are slightly different, which 
 could be one of the sources of disagreement between the QA-frame symmetrized waveforms. However, the similarity of the trends 
 of the match vs $f_{max}$ for all three systems indicate that the above results should hold within these uncertainties.    


\subsection{Inertial frame symmetrized waveform analysis}~\label{sec:inertial_symm_wvfm_results}

\begin{figure}
	\includegraphics[width=0.48\textwidth,keepaspectratio]{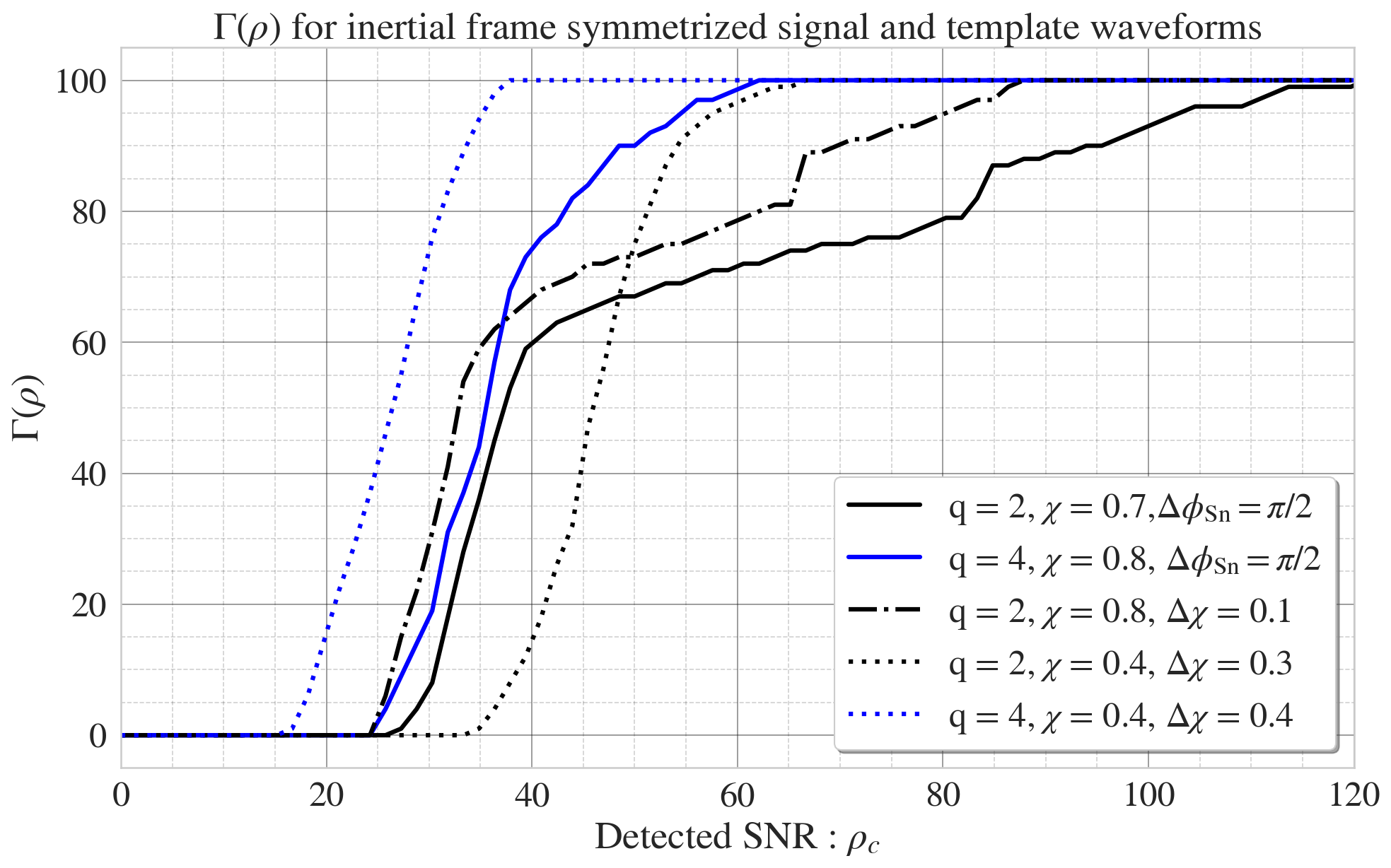}
	\caption{$\Gamma (\rho)$ from matching waveforms symmetrized in the QA-frame and then transformed back to the inertial frame.  The systems q2a07p90 (solid-black), q2a08p0 (dashed-dotted -black), q2a04p0 (dashed-black) are matched with the q2a07p0 proxy template. The q4a08p90 (solid-blue) and q4a04p0 (dashed-blue) systems are matched with q4a08p0 template. The legend shows the mass-ratio and spin of signal and the difference in the relevant parameter with respect to the template.}~\label{fig:s12b_symm_vs_symm}
\end{figure}

We now 
transform the symmetrized QA frame waveforms to the inertial frame using Eq.~(\ref{eq:wigner_rotation}) and the 
corresponding $(\alpha, \beta, \epsilon)$ angles for each system. This is similar to how current waveform models construct the 
precessing waveforms in the inertial frame, i.e., they transform a model for the corresponding aligned-spin QA frame waveform 
to the inertial frame using a model for the precession Euler angles. Using these waveforms, we perform the same analysis as in 
Sec.~\ref{sec:full_wvfm_results} and plot the $\Gamma(\rho)$ quantity in Fig.~\ref{fig:s12b_symm_vs_symm}. Note, that for the 
$q=1$ system, the symmetrized waveform matches will be the same as presented in Fig.~\ref{fig:s123b_qa_symm} and we will 
not discuss that system here. 

We consider first the two $\phi_{\rm{Sn}} =  \pi/2$ configurations for the $q=2$ and $q=4$ systems. 
Between the symmetrized q2a07 systems, the distinguishability SNR is $25 \lesssim \rho_c \lesssim 115$. Between the 
symmetrized q4a08 systems, the distinguishability SNR is $25 \lesssim \rho_c \lesssim 60$. In both cases, this is 
significantly higher than for the waveforms with mode-asymmetry included, and the range is either side of the value for the symmetrized QA-frame waveforms. 
In particular, we see that the presence of asymmetries makes the q2a07 cases distinguishable at SNRs as low as 20, and the 
q4a08 cases distinguishable at SNRs as low as 10, while, if the asymmetries did not exist, they would not be distinguishable for
SNRs lower than $\sim$30. 

If we now consider the distinguishability between configurations with different spin magnitudes, comparing Figs.~\ref{fig:s123b_fullvfull} and
\ref{fig:s12b_symm_vs_symm}, we see a similar effect. Although for $q=2$ configurations the $\rho_c$ for different spin magnitudes show an 
overall increase, a few of the signals with a $\chi_p$ difference of 0.1 are now easier to distinguish than a $\chi_p$ difference of 0.3. The most 
pronounced effect is for $q=4$ configurations, where the spin difference of 0.4 is now easier to distinguish than the rotation of the spin.

\begin{figure}
	\includegraphics[width=0.48\textwidth,keepaspectratio]{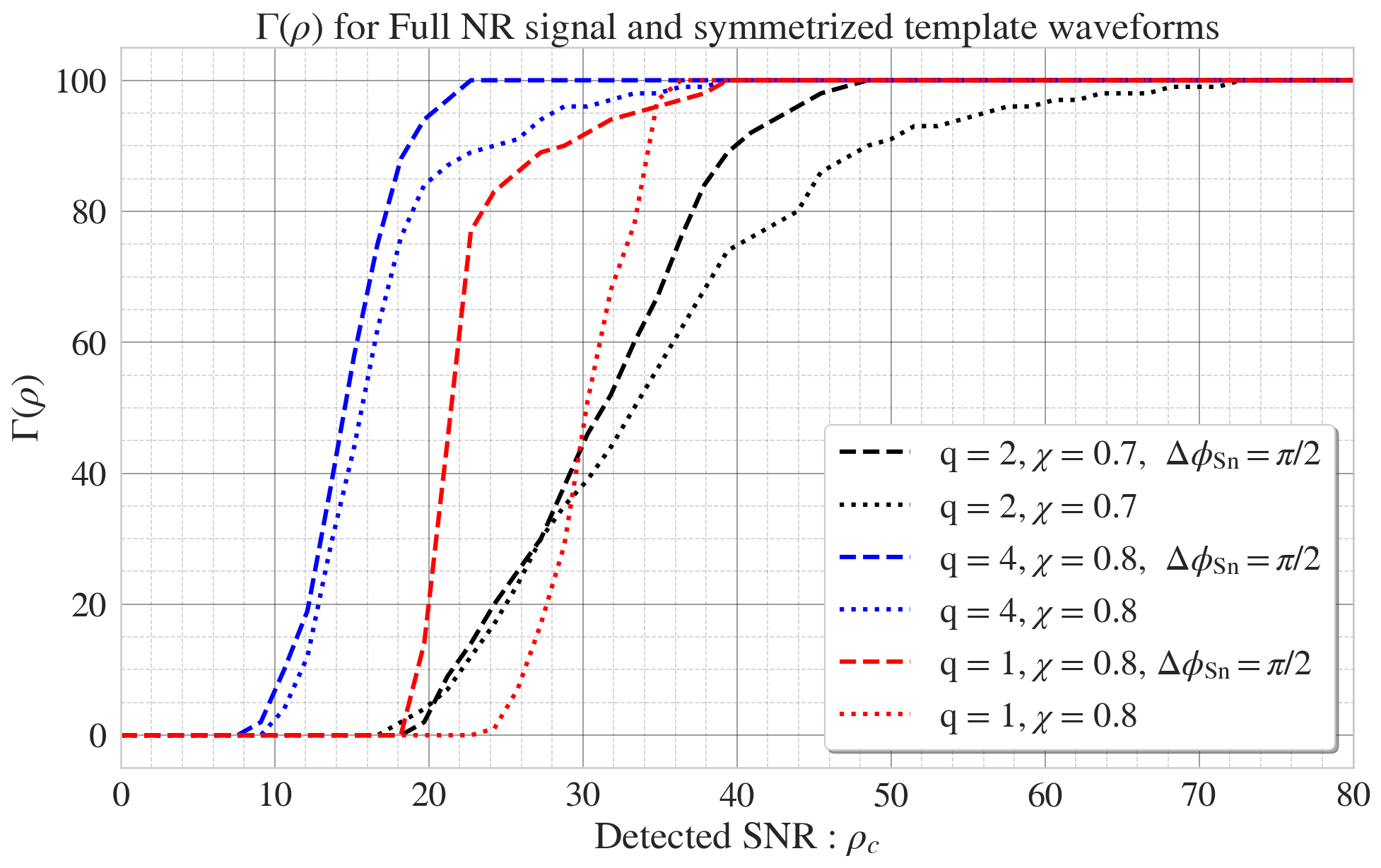}
\caption{ $\Gamma (\rho)$ from matching signal waveforms with both precession and mode-asymmetry against symmetrized template waveforms.  Here, the full 
waveform $\phi_{\rm{Sn}} = \pi/2$ signal and $\phi_{\rm{Sn}} = 0$ signal as seen by the symmetrized $\phi_{\rm{Sn}} = 0$ 
template are shown by the dashed and dotted lines respectively. The legend shows the mass-ratio and spin of signal and the difference in the relevant parameter with respect to the template. Where no parameter difference is mentioned, it shows the result for the signal with both precession and mode-asymmetry against its symmetrized self.}~\label{fig:s123ab_full_vs_symm}
\end{figure} 

Neither of the previous analyses reflects the scenario of current GW measurements, where the signals correspond to ``full'' waveforms, i.e, signals with both precession and mode-asymmetry,
and they are analysed using models that correspond approximately to the symmetrized waveforms of the previous analysis. In order to estimate the impact of using symmetrized models in analysis, 
Fig.~\ref{fig:s123ab_full_vs_symm} shows $\Gamma(\rho)$ for the $\phi_{\rm{Sn}} = 0, \pi/2$ full waveform signals matched 
against the symmerized inertial frame $\phi_{\rm{Sn}} = 0$ templates for the $q=1$, $q=2$ and $q=4$ simulations. 
We observe that for $\phi_{\rm{Sn}} = \pi/2$, it is generally easier for the symmetrized template to distinguish the signal as 
compared to the full waveform templates. This effect is very strong for the $q=1$ super-kick cases where the distinguishability 
SNR reduces by almost 20 for all signals. We also see that removing the mode-asymmetric content leads to large mis-matches 
between waveforms of the same systems causing the full $\phi_{\rm{Sn}}=0$ signal to be distinguishable from the 
symmetrized $\phi_{\rm{Sn}}=0$ template at moderate $(10 < \rho_c < 40)$ SNRs for all mass ratios. All these results indicate that 
the absence of mode asymmetries in current models will lead to measurement biases in these systems. We expect that even
for comparable-mass systems, if the total mass is high ($> 100\,M_\odot$) and the in-plane spins are high, systematic errors
are likely to be significant. 

In Tab.~\ref{Tab:system_match_all}, we list the SNR averaged match values over all the signal $(\theta, \phi, \psi)$ values to provide one 
single number for the distinguishability of the signal. We can see that when both signal and templates are symmetrized, for all systems, the 
agreement between the waveforms increases leading to larger distinguishability SNR. When symmetrized waveform templates are matched 
with full waveform signals, we see an overall decrease in the distinguishability SNR. Even when both the signal and template systems are the 
same, with symmetrized templates, $\rho_c$ is comparable to that of $\phi_{\rm{Sn}} \pm \pi/2$ results.

	\begin{table}[t]
	\resizebox{0.48 \textwidth}{!}{\begin{tabular}{|c|c|c|c|c|c|}
	\hline 
	Template & Signal & Signal Effects & Template Effects & Average Match & $\rho_c$  \\
	\hline
	 & $\phi_{\rm{Sn}} +\pi/2$ & Full & Full & 0.9983 & 60 \\
	 & $\phi_{\rm{Sn}} +\pi/2$ & Symmetrized & Symmetrized & 0.9991 & 83 \\
	 & $\phi_{\rm{Sn}} +\pi/2$ & Full & Symmetrized & 0.9954 & 36 \\
	$q=2,\chi=0.7,\phi_{\rm{Sn}}=0$ & $\phi_{\rm{Sn}} + 0$ & Full & Symmetrized & 0.9969& 44 \\
	``q2a07p0'' & $\chi+0.1$ & Full & Full & 0.9983 & 60 \\
	& $\chi+0.1$ & Symmetrized & Symmetrized & 0.9986 & 65 \\
	 &  $\chi-0.3$ & Full & Full & 0.9952 & 36 \\
	 &  $\chi-0.3$ & Symmetrized & Symmetrized & 0.9969 & 44 \\
	\hline
	\hline
	 &  $\phi_{\rm{Sn}} +\pi/2$ & Full & Full & 0.9811 & 18 \\
	 &  $\phi_{\rm{Sn}} +\pi/2$ & Symmetrized & Symmetrized & 0.9935 & 30 \\
	$q=4,\chi=0.8,\phi_{\rm{Sn}}=0$ &  $\phi_{\rm{Sn}} +\pi/2$ & Full & Symmetrized & 0.9737 & 15 \\
	 ``q4a08p0'' &  $\phi_{\rm{Sn}} + 0$ & Full & Symmetrized & 0.9785 & 16 \\
	 & $\chi-0.4$ & Full & Full & 0.9936 & 30 \\
	& $\chi-0.4$ & Symmetrized & Symmetrized & 0.9942 & 32 \\
	\hline
	\hline
	$q=1,\chi=0.8,\phi_{\rm{Sn}}=0$ & $\phi_{\rm{Sn}} +\pi/2$ & Full & Full & 0.9981 & 57 \\
	``q1a08p0$_{\rm {sk}}$'' & $\phi_{\rm{Sn}} +\pi/2$ & Full & Symmetrized & 0.9882 & 22 \\
	 & $\phi_{\rm{Sn}} + 0$ & Full & Symmetrized & 0.9934 & 30 \\
	\hline
	\end{tabular}}
	\caption{The SNR averaged match, Eq.~(\ref{eq:snr_aveg_match}), over all the $(\theta, \phi, \psi)$ values for the systems considered 
	in this study. From left to right, the columns state the template waveform configuration, difference in the relevant parameter between the template and signal waveform, signal effects (full waveform or symmetrized), teamplate effects (full waveform or symmetrized), average match value and corresponding SNR respectively, using 
Eq.~(\ref{eq:det_snr}). See text for further discussion.}~\label{Tab:system_match_all}
\end{table}

\section{Conclusions}~\label{sec:conclusions}

We have investigated when changes in the in-plane spin direction of binary-black-hole systems, the effects of which are 
not included in current Phenom and EOB models, will be distinguishable in GW measurements. 
To do that, we use a set of NR simulations obtained from the BAM code (see Tab.~\ref{Tab:nr_waveforms_list_ch4}). 
We quantify the distinguishability of systems with different choices of in-plane spin direction $\phi_{\rm{Sn}}$ by calculating matches
between them. This approach allows us to estimate the SNR at which the signals will be distinguishable. Our study is restricted to 
a small number of configurations at mass ratios $q=1, 2, 4$, and large in-plane spin magnitudes of 0.7 and 0.8, with two configurations with moderate in-plane spin of 0.4. All of our calculations are performed on systems with total mass $100\,M_\odot$. 

Changes in $\phi_{\rm{Sn}}$ have several effects on the binary dynamics and the waveform. One effect that we discuss in detail is 
the asymmetry between the $\pm m$ waveform modes. Another is small changes in the phasing of the binary, and in the merger
and ringdown signal. By removing asymmetry and/or precession effects from our waveforms, we show that all of these effects 
contribute to the waveform variations between different choices of  $\phi_{\rm{Sn}}$.
When mode-asymmetries are muted, the distinguishability SNR $\rho_c$ for \emph{all} the systems (different $\phi_{\rm{Sn}}$ and 
different $\chi_p$) show a marked increase across the ($\theta, \phi$) space (see Fig.~\ref{fig:s12b_symm_vs_symm}). 
Disregarding mode asymmetries increases $\rho_c$ by factors of $\sim 1.5 - 1.9$ between systems of different $\phi_{\rm{Sn}}$,
indicating that this is a significant feature of these waveforms.  

Our main results are shown in Sec.~\ref{sec:full_wvfm_results}, and show that for large in-plane spins, variations in $\phi_{\rm{Sn}}$
will be distinguishable at moderate SNRs. More importantly, these effects will influence measurements at SNRs comparable to those
at which in-plane spin magnitudes become measurable. For example, in the $q=2$ systems we considered, a change in spin 
magnitude of 0.3 will be distinguishable at a comparable SNR to a change in spin direction of $\pi/2$. This effect will be reduced
for smaller spins, but so will our ability to measure the spin magnitude. Precession effects and in-plane spin magnitude, typically 
captured by the parameter $\chi_p$, have not yet been identified in individual observations~\cite{Fairhurst:2019srr}. Our results 
suggest that when they are, the absence of in-plane spin direction effects in the modelling could lead to significant parameter
biases. We plan to study the impact on parameter estimation in future work. 

There are a number of questions that require further work. We have limited ourselves to small number of configurations, and to one
choice of total mass. We have also neglected the effect of $\ell > 2$ modes, which also impact parameter estimation for systems with 
mass ratios of $q \geq 2$~\cite{Kalaghatgi:2019log}. The impact of changes in $\phi_{\rm{Sn}}$, and the importance of mode
asymmetries, also needs to be studied for systems with lower masses, where the inspiral contributes more power to the waveform, 
with mode-asymmetric effects being weaker but with a larger number of precessional cycles. However, in order to fully understand the 
importance of these physical effects,
we require models that include them, which can then be used in parameter-estimation studies. This work has provided strong
evidence that these effects must be taken into account in order to make unbiassed physical measurements from GW 
observations, and therefore already provide a strong motivation for such modelling. This has already been done for the 
surrogate models described in Refs.~\cite{Blackman:2017pcm,Varma:2019csw}. Since these models are valid only for high-mass
systems and a limited range of mass ratios, it would be advantageous to be extended to other classes of model.

\section{Acknowledgements}

We thank Frank Ohme and Sebastian Khan for useful discussions. We thank Edward Fauchon-Jones, Eleanor Hamilton, Charlie Hoy and Dave Yeeles for their help in performing the comparison cases NR simulations.

This work was supported by Science and Technology Facilities Council (STFC) grant ST/L000962/1, European
Research Council Consolidator Grant 647839.
We are grateful for computational resources provided by Cardiff University, and funded by an
STFC grant supporting UK Involvement in the Operation of Advanced LIGO. 

Numerical simulations were performed on the DiRAC@Durham facility managed by the Institute for Computational Cosmology 
on behalf of the STFC DiRAC HPC Facility (www.dirac.ac.uk). The equipment was funded by BEIS capital funding via STFC 
capital grants ST/P002293/1, ST/R002371/1 and ST/S002502/1, Durham University and STFC operations grant ST/R000832/1. 
DiRAC is part of the National e-Infrastructure.


\appendix
\section{Initial data generation}

~\label{sec:ini_dat_gen}
For this study, we required singe-spin precessing NR waveforms with user specified $(\theta_{SL}, \phi_{\rm{Sn}})$ at a given reference frequency $M \omega_{orb}$. Over the course of inspiral, the spin vectors of a precessing system oscillate about a mean value with the oscillation frequency increasing as system nears merger~\cite{Schmidt:2014iyl}. An iterative method was required to ensure the required spin direction at the given reference frequency. The code used for solving the PN equations was one which was used for BAM NR waveforms as used in~\cite{Hannam:2010ec, Schmidt:2012rh, Husa:2015iqa}. The method developed for initial data generation is as below. The PN evolution is started in the $\vec{J}$ aligned to $\hat{z}$ frame with $\vec{L}$ being the Newtonian angular momentum. 

The angle between the spin vector and angular momentum vector, $(\theta_{SL})$, varies not more than$\sim 1^{\circ}$ over the inspiral phase. Hence, once $(\theta_{SL})$ is specified, further iteration is not required To obtain the required $\phi_{\rm{Sn}}$, the algorithm goes through the following steps:

\textbf{Step 1}:	

This step consists of two iterations. 

\textbf{Iteration 1}:
Initially, both the BHs are placed along the x-axis with a given separation, the orbital plane is the $x$-$y$ plane, and the initial spin ($\mathbf{S}_{ini}$) parallel to $\hat{n}$. The spin vector is then rotated to obtain the required $\theta_{SL}$ and the PN evolution code is run until $M \omega_{orb}$ is reached. We record the time when the specified orbital frequency is reached ($t_{0}$), the value of $\phi_{\rm{Sn}}(t)$ at $t_0$ [$\phi_{\rm{Sn}}(t_0)$], the closest time to $t_{0}$ at which $\phi_{\rm{Sn}}(t) = \phi_{\rm{Sn}}^{target}$, which is denoted $t_1$, and finally the relative frequency error ($\omega_{err}$) between the orbital frequencies at $t_0$ and $t_1$. If, at this iteration, $\phi_{\rm{Sn}}$ at $t_0$ is not $\phi_{\rm{Sn}}^{target}$ or if $\omega_{err}$ is larger than a pre-specified threshold ($\omega_{err}^{F}$), the value of $\phi_{\rm{Sn}}^{t_1}$ is recorded; we call this $\phi_{\rm{Sn}}^{1}$. Each iteration hence also stores the value $\phi_{\rm{Sn}}^{t_0, i}$. For these simulations, we use $\omega_{err}^{F} = 1\%$
	
\textbf{Iteration 2}:
During iteration 2, $\mathbf{S}_{ini}$ is rotated to obtain the required $\theta_{SL}$ and then further rotated by ($- \phi_{\rm{Sn}}^{1}$) about the $z$-axis, and then the PN solver is again run. If the conditions specified in Iteration 1 are met ($\omega_{err} <\omega_{err}^{F}$  \& $\phi_{\rm{Sn}}^{t_1}$ = $\phi_{\rm{Sn}}^{target}$) then the parameters at $t_1$ are recorded. If not, we would ideally simply repeat the process. However, since
$\phi_{\rm{Sn}}$ changes on the (rapid) orbital timescale, the value of $\phi_{\rm{Sn}}$ at the NR reference frequency is very sensitive to the choice
at the beginning of the PN evolution, and so this procedure is not well-conditioned to fine-tune $\phi_{\rm{Sn}}$. 
We instead proceed to Step Two, and store the value of $\phi_{\rm{Sn}}^{t_1}$ of this iteration as $\phi_{\rm{Sn}}^{2}$. 
	
\textbf{Step 2}:
Depending on the parameters, this step can consist of one or multiple iterations. For each iteration, $\mathbf{S}_{ini}$ is rotated to obtain the required $\theta_{SL}$ and then by the specified $- \phi_{rot}$ about $\hat{z}$.  

\textbf{Iteration 3}:
For each iteration hence, we define a angle correction parameter, $\phi_{corr}$. $\omega_{err} - \omega_{err}^{F}$ gives an idea of how close we are to the required initial parameters and value of $\phi_{corr}$ is based on that. If, $\omega_{err} - \omega_{err}^{F} > \frac{1}{2} \omega_{err}^{F} $, then $\phi_{corr} = 10^{o}$, else $\phi_{corr} = 5^{o}$ and then $\phi_{rot} = \phi_{\rm{Sn}}^{2} + \phi_{corr}$. Using these angles, the spin is rotated and PN solver is run. Again, the value of $\phi_{\rm{Sn}}^{t_1}$ of this iteration as $\phi_{\rm{Sn}}^{3}$.

\textbf{Iteration \textit{n}$> 3$ }:
First, we check if $\phi_{\rm{Sn}}^{3} > \phi_{\rm{Sn}}^{2}$. If so, the initial spin is being rotated in the wrong direction and for each subsequent iteration, $\phi_{rot} = \phi_{\rm{Sn}}^{2} - (n-3) \times \phi_{corr}$, if not, $\phi_{rot} = \phi_{\rm{Sn}}^{2} + (n-2) \times \phi_{corr}$. Thus, we brute force the initial direction of $\mathbf{S}_{ini}$ until the required direction of $\mathbf{S}$ is obtained at the reference frequency. 

To apply this procedure with a higher tolerance, one should reduce $\phi_{corr}$ in subsequent iterations. For the simulations produced here,
no more than two or three iterations in Step Two were required. 

\bibliographystyle{ieeetr}
\bibliography{spin_paper.bib}
\end{document}

%% file: misc-definitions.tex
\usepackage{inputenc}
\usepackage{pgfplots}
\usepackage{tikz}
\usetikzlibrary{shapes,arrows}
\usetikzlibrary{positioning}
\usetikzlibrary{backgrounds}

\definecolor{lightblue}{rgb}{.82,.88,0.95}
\definecolor{lightred}{rgb}{0.95,.86,0.86}
\definecolor{yellow}{rgb}{0.95,0.95,0.86}
\definecolor{green}{rgb}{.90,1,0.95}
\definecolor{lightpurple}{rgb}{.95,0.85,0.95}

\tikzset{
    vertex/.style = {
        circle,
        fill            = black,
        outer sep = 2pt,
        inner sep = 1pt,
    }
}

\def\gw#1{gravitational wave#1}

\def\rm#1{\mathrm{#1}}

\def\imr#1{inspiral, merger and ringdown#1
  (IMR#1)\gdef\imr{IMR}}
\def\lal#1{LIGO Analysis Library#1
  (LAL#1)\gdef\lal{LAL}}
\def\nrda#1{\nr{} Data Analysis#1
  (NRDA#1)\gdef\nrda{NRDA}}
\def\tt#1{\textit{transverse--traceless}#1
  (TT#1)\gdef\tt{TT}}
\def\et#1{Einstein Telescope#1
  (ET#1)\gdef\et{ET}}
\def\ego#1{European Gravitational Observatory#1
  (EGO#1)\gdef\ego{EGO}}
\def\elisa#1{Evolved Laser Interferometer Space Antenna#1
  (eLISA#1)\gdef\elisa{eLISA}}
\def\ligo#1{Laser Interferometer Gravitational Wave Observatory#1
  (LIGO#1)\gdef\ligo{LIGO}}

\def\aligo#1{Advanced LIGO#1
  (aLIGO#1)\gdef\aligo{aLIGO}}
\def\snr#1{signal-to-noise ratio#1
  (SNR#1)\gdef\snr{SNR}}
\def\psd#1{power spectral density#1
  (PSD#1)\gdef\psd{PSD}}
\def\rom#1{reduced order model#1
  (ROM#1)\gdef\rom{ROM}}
\def\gatech#1{Georgia Institute of Technology#1
  (GaTech#1)\gdef\gatech{GaTech}}
\def\ffi#1{Fixed-Frequrncy Integration#1
  (FFI#1)\gdef\ffi{FFI}}
\def\sxs#1{Simulating Extreme Spacetimes#1
  (SXS#1)\gdef\sxs{SXS}}
\def\bam#1{Bifunctional Adaptive Mesh#1
  (BAM#1)\gdef\bam{BAM}}
\def\adm#1{Arnowitt-Deser-Misner
	(ADM#1)\gdef\adm{ADM}}
\def\frmse#1{Fractional Root-Mean Square Error
	(FRMSE)\gdef\frmse{FRMSE}}

\def\bh#1{black hole#1
 (BH#1)\gdef\bh{BH}}

\def\qnm#1{Quasi-Normal Mode#1
  (QNM#1)\gdef\qnm{QNM}}
\def\eob#1{Effective One Body#1
  (EOB#1)\gdef\eob{EOB}}
\def\gw#1{gravitational-wave#1}
\def\gw#1{Gravitational wave#1
 (GW#1)\gdef\gw{GW}}

\def\spa#1{stationary phase approximation#1
 (SPA#1)\gdef\spa{SPA}}
\def\pn#1{Post-Newtonian#1
	(PN#1)\gdef\pn{PN}}
\def\pnl#1{post-Newtonian-like#1
  (PN-like#1)\gdef\pnl{PN-like}}

\def\nr{Numerical Relativity
 (NR)\gdef\nr{NR}}

\def\rd{ringdown}
\def\imr{inspiral, merger and \rd{}}

\def\pca#1{principle component analysis#1
  (PCA#1)\gdef\pca{PCA}}
\def\svd#1{Singular Value Decomposition#1
  (SVD#1)\gdef\svd{SVD}}

\newcommand\hidetosubmit[1]{}
\newcommand\optional[1]{}
\newcommand\ForInternalReference[1]{}